\documentclass[english,iop,apj]{emulateapj}
\usepackage[T1]{fontenc}
\usepackage{verbatim}
\usepackage{graphicx}
\usepackage{subfigure}
\usepackage{tabularx}
\usepackage{amsmath}
\usepackage{natbib}
\usepackage{dsfont}
\usepackage{color}
\usepackage[english]{babel}

\makeatletter  

\slugcomment{}

\shorttitle{Dust filtering in the early solar system}
\shortauthors{Haugb{\o}lle et al.}

\usepackage{txfonts}
\usepackage{cleveref}

\newcommand{\fdrag}{\mathbf{f}_{\rm drag}}

\newcommand{\vvg}{\mathbf{u}}
\newcommand{\vvd}{\mathbf{v}}

\newcommand{\St}{S\! t}

\usepackage{color,ulem,soul}
\definecolor{dark-red}{rgb}{0.75, 0.00, 0.00}
\definecolor{hlcolor}{rgb}{1.00, 0.90, 0.85}\sethlcolor{hlcolor}
\definecolor{gray}{rgb}{0.7, 0.7, 0.7}
\definecolor{ctroels}{rgb}{0.65, 0.00, 0.65}
\definecolor{crevision}{rgb}{0.4, 0.0, 0.4}

\newcommand{\rev}[1]{#1}

\makeatother 

\begin{document}

\title{Probing the protosolar disk using dust filtering at gaps in the early Solar System}
\author{Troels Haugb{\o}lle$^{1,\dagger}$}
\author{Philipp Weber$^2$}
\author{Daniel P.~Wielandt$^3$}
\author{Pablo Ben{\'i}tez-Llambay$^2$}
\author{Martin Bizzarro$^4$}
\author{Oliver Gressel$^{2, 5}$}
\author{Martin E.~Pessah$^2$}
\affil{$^1$Niels Bohr Institute and Centre for Star and Planet Formation, University of Copenhagen,
{\O}ster Voldgade 5, DK-1350 Copenhagen K, Denmark}
\affil{$^2$Niels Bohr International Academy, Niels Bohr Institute, University of Copenhagen,
Blegdamsvej 17, DK-2100 Copenhagen {\O}, Denmark}
\affil{$^3$Quadlab and Natural History Museum of Denmark, University of Copenhagen, {\O}ster Voldgade 5, DK-1350 Copenhagen K, Denmark}
\affil{$^4$Centre for Star and Planet Formation and Natural History Museum of Denmark, University of Copenhagen,
{\O}ster Voldgade 5, DK-1350 Copenhagen K, Denmark}
\affil{$^5$Leibniz-Institut f{\"u}r Astrophysik Potsdam (AIP), An der Sternwarte 16, 14482 Potsdam, Germany}
\email{$^\dagger$haugboel@nbi.ku.dk}

\date{\today}

\begin{abstract}
Jupiter and Saturn formed early, before the gas disk dispersed.
The presence of gap-opening planets affects the dynamics of the gas and embedded solids
and halts the inward drift of grains above a certain size. 
A drift barrier can explain the absence of calcium aluminium rich inclusions (CAIs)
in chondrites originating from parent bodies that accreted in the inner solar system.
Employing an interdisciplinary approach, we use a $\mu$-X-Ray-fluorescence scanner to search for large CAIs and a
scanning electron microscope to search for small CAIs in the ordinary chondrite NWA 5697.
We carry out long-term, two-dimensional simulations including gas, dust, and planets
to characterize the transport of grains within the viscous $\alpha$-disk framework exploring the scenarios
of a stand-alone Jupiter, Jupiter and Saturn \textit{in situ}, or Jupiter and Saturn in a 3:2 resonance.
In each case, we find a critical grain size above which drift is halted as a function of the physical conditions in the disk.
From the laboratory search we find four CAIs with a largest size of $\approx$200$\,\mu$m. \rev{Combining models and data,
we provide an estimate for the upper limit of the $\alpha$-viscosity and the surface density at the location of Jupiter, using
reasonable assumptions about the stellar accretion rate during inward transport of CAIs, and assuming angular momentum
transport to happen exclusively through viscous effects.}
Moreover, we find that the compound gap structure in the presence of Saturn \rev{in a 3:2 resonance} favors inward transport of \rev{grains larger
than CAIs currently detected in ordinary chondrites}.
\end{abstract}


\maketitle

\section{Introduction} \label{sec:intro}

In the young solar system, planets formed out of the protosolar accretion disk that contained a mixture of gas, dust, and ices.
While most material was either accreted by the young Sun or recycled to the molecular cloud through outflows, a small
percentage became the building blocks for the planets or was left over in a debris disk that evolved into the asteroid and
Kuiper belts we observe today.
The evolution, dynamics, and concentration of dust in the protosolar accretion disk are crucial ingredients in
understanding how the solar system architecture was created. The dynamics of dust grains is dictated through
friction forces by the relative velocity to the gas, and small dust grains will follow the gas flow, while larger grains decouple from the
gas. In general, in a protoplanetary disk the gas pressure is monotonically increasing toward the center of the disk, resulting in an
outward pressure gradient force canceling part of the gravitational pull and a slightly sub-Keplerian gas rotational velocity. In contrast,
the nearly collisionless dust would rotate with Keplerian velocities if interactions were absent. But friction forces between
gas and dust result in a constant headwind on the dust grains, a drain of their angular momentum, and a rapid in-spiraling motion for
millimeter- to meter-sized objects on a time-scale of thousands of years, leading to the so-called drift barrier for smaller bodies and pebbles
\citep{1977MNRAS.180...57W,1980ApJ...241..425G}.

In general, the formation of gas giants has to happen while significant amounts of gas are still present in the accretion disk, implying that friction forces between gas and dust are still important at this stage. Specific to the solar system, the core of proto-Jupiter probably had enough gravitational
pull to exert a large tidal perturbation on the disk and induce strong trailing shocks. Through transport of angular momentum and accretion it would
deplete its orbit of material and thereby perturb the structure of the disk and create a gap \citep{1979MNRAS.186..799L,1980ApJ...241..425G}
already at early times. In a viscously evolving disk, the gap cannot be completely devoid of gas, and there remains an accretion flow
through it \citep{1986ApJ...309..846L,1996ApJ...467L..77A}. Small grains remain well coupled to the gas and will be transported through
the gap together with the gas. Friction forces between gas and larger grains will concentrate the grains in a local maximum in the gas pressure
that develops just outside the gap and inhibit their passage through the gap \citep{1972fpp..conf..211W}.
The formation of a large planet will therefore both change the gas structure and act as a trap for larger dust grains, with a size-dependent efficiency.

Evidence from primitive meteorites coming from parent bodies that have remained unprocessed since the formation of the solar system gives a
unique window into the processes operating in the early stages of planet formation.
Very precise measurements can be done in the laboratory, but the material collected from meteorites falling on Earth has an
unknown dynamical history and comes with large and unknown sampling biases. The full potential of such measurements can
therefore only be realized when put into a global context through the use of astrophysical models that provide the larger picture
wherein the results can be interpreted. Primitive meteorites are mostly composed of quasi-spherical glassy chondrules formed by
transient heating events in the protosolar disk; refractory inclusions, such as calcium aluminum rich inclusions (CAIs), formed
by condensation from a hot metal-rich gas; and fine-grained matrix material.

From detailed numerical modeling \citep{2016ApJ...826...22K} and simple estimates of the mixing time-scales of the pre-stellar material,
there is strong evidence that the very early evolution was dominated by accretion from a well-mixed inner part of the pre-stellar core.
At later stages, past the first 100$\,$kyr, corresponding to material originating from outside the inner $\sim$4000$\,$AU of the pre-stellar core,
matter may have been isotopically distinct. 

This is supported by the isotopic composition of nucleosynthetic tracers such as $^{54}$Cr in chondrites and their components,
suggesting the existence of three distinct disk reservoirs \rev{\citep{2011E&PSL.311...93W}}. The ordinary and enstatite chondrites, characteristic
of the inner solar system based on their water content \citep{2012AREPS..40..251M}, are depleted in $^{54}$Cr relative to
Earth \rev{\citep{2007ApJ...655.1179T,2010GeCoA..74.1122Q}}.
The metal-rich chondrites, which are thought to have formed beyond the orbit of Saturn, are systematically enriched in $^{54}$Cr, whereas the rest of the carbonaceous chondrites, which are predicted to have formed in between Jupiter and Saturn, contain a mixture of material with isotopic signatures characteristic of the inner and outer solar system \rev{\citep{2016GeCoA.191..118O,2016E&PSL.454..293B,2017LPICo1975.2008B,2017GeCoA.208....1V,2018ApJ...854..164S}}.

These isotopic variances are also recorded in
the titanium isotopes of the bulk chondritic material \citep{2009Sci...324..374T}. U--Pb dating of individual chondrules shows a large age span of up
to 4 Myr for individual chondrules inside a single chondrite \citep{2017SciA....3E0407B}. The chondrule ages taken together with the isotope
measurements for individual chondrules and bulk chondritic material support the idea that distinct reservoirs already existed very early in the solar system.

From absolute U--Pb dating, it is well established that CAIs formed very early at the
birth of the solar system \citep{2012Sci...338..651C}. Moreover, the CAI factory must have operated for a short time interval
of a few thousands of years based on the $^{26}$Al--$^{26}$Mg systematics of individual CAIs \rev{\citep{2011ApJ...735L..37L,2008E&PSL.272..353J}.
Refractory inclusions formed very close to the proto-Sun, where the necessary high densities and temperatures required for condensation
and high levels of irradiation recorded by the former presence of $^{10}$Be in the gas reservoir were present
\citep{2000Sci...289.1334M,2012ApJ...748L..25W,2013ApJ...763L..33G}.}
But, remarkably, CAIs are almost exclusively found today in carbonaceous chondrites that are thought to have accreted in the outer solar system.
The mass abundance of CAIs range from subpercentage up to almost 10\% for certain CK, CV, and CO chondrites
\citep{2008M&PS...43.1879H}. In contrast, only a handful of CAIs are known in inner solar system ordinary chondrites (OCs),
and the titanium isotopic ratio in bulk material from OCs only allows for a relative bulk CAI mass fraction of
less than 0.1\%. This low abundance region also includes enstatite chondrites, as well as Mars and Earth \citep{2009Sci...324..374T}. 

The dichotomy in the abundance of CAIs can be understood in a scenario where CAIs were accreted by the proto-Sun
right after their production, or sent out in outflows that propelled them either out into the interstellar medium or far out
into the protostellar envelope \rev{\citep{1996Sci...271.1545S,2016MNRAS.462.1137L,2017LPICo1975.2025H,2018ApJ...854..164S}} from where they were
transported back, as part of the general gas infall, toward the outer solar system. \rev{The vertical transport in the disk needed to embed CAIs
in the outflow are in accordance with the cooling times of 1--10 K / hr found in laboratory experiments to be necessary for creating refractory condensates
with properties similar to Type B CAIs \citep{2002GeCoA..66..521R}. This could be achieved by formation in an optically thick environment,
with subsequent transport away from the region, but has to happen faster than on orbital or viscous timescales.
Once CAIs are reinserted into the outer disk as part of the envelope inflow, they} start to drift inward
until stalling at local extrema in the gas pressure because of gaps in the gas distribution opened by the gas giants.
Alternatively, it has been proposed that CAIs were transported
outward in the disk through turbulent diffusion, as part of meridional flows \citep{2003Icar..166..385C,2010ApJ...719.1633H},
\rev{or in a viscous spreading disk \citep{2012M&PS...47...99Y,2019E&PSL.511...44N}.
While also resulting in transport of CAIs to the outer part of the protosolar disk, these mechanisms confine the CAIs to the
protosolar disk, and the viscous transport at much lower speeds than through outflows would result in lower CAI cooling rates.
In such a model, the CAIs are confined to the disk, and this implies significantly shorter drift timescales
and correspondingly shorter timescales for forming the gas giants and establishing dust traps such that CAIs do not reenter the inner solar system
\citep[see, e.g.~][that requires formation the core of Jupiter in 0.6 Myr]{2018ApJS..238...11D}. Differentiated planetesimals
are suggested to have accreted within a few $\times 10^5$ years of CAI formation
\citep{2011ApJ...740L..22S,2014Sci...344.1150K,2015E&PSL.420...45S,2016GeCoA.176..295L} and outward diffusion of CAIs through the
protosolar disk would result in the addition of refractory material within early formed planetesimals. However, the $^{54}$Cr--$^{50}$Ti systematics of inner solar system bodies allow for very little CAI material in their precursors, implying that outward transport through the disk may not have been the dominant mechanism \citep{2016GeCoA.191..118O,2017ASSL..445..161B}.}

\begin{figure}
        \centering
        \includegraphics[width=0.95\columnwidth]{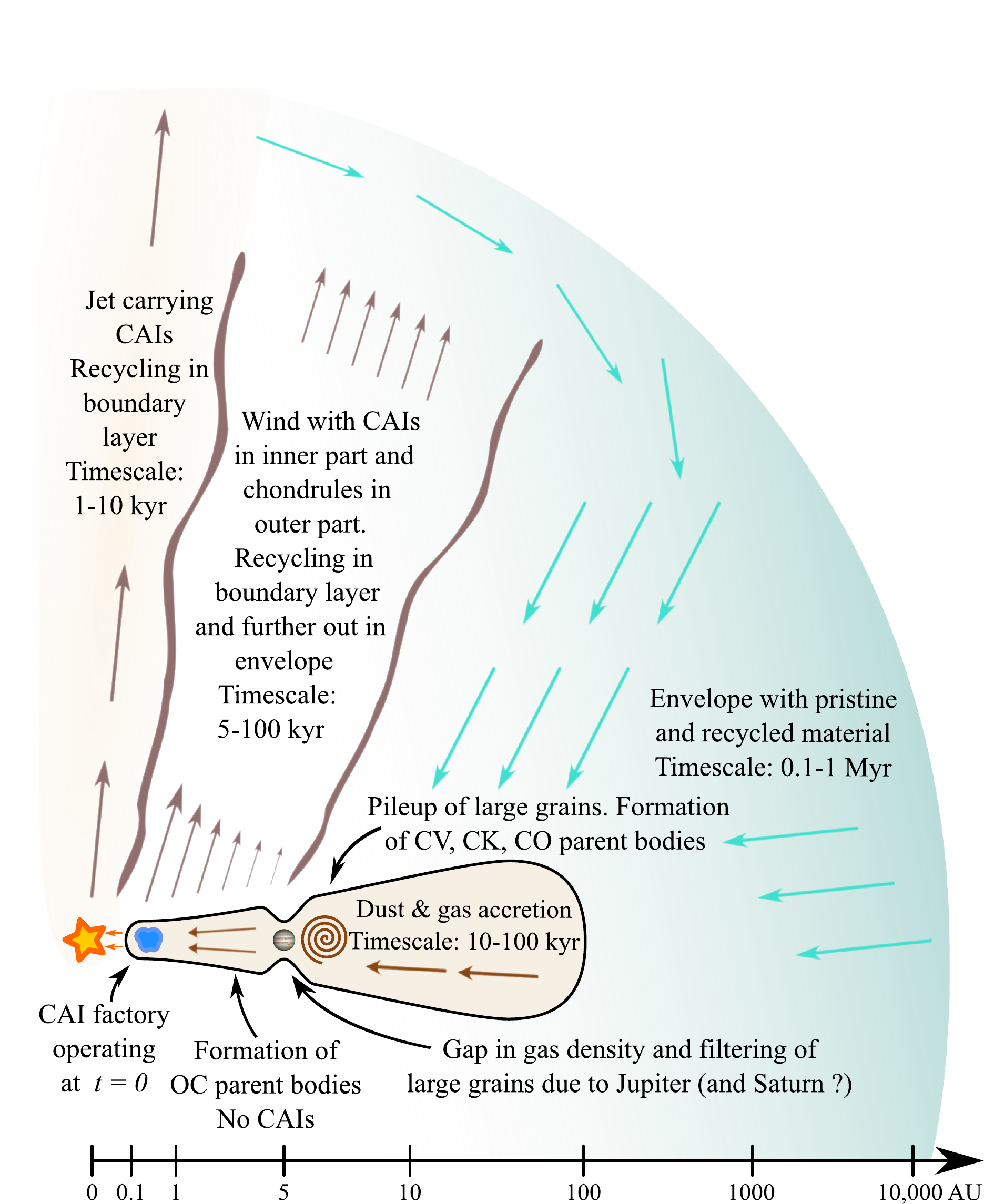}
        \caption{Sketch of the protosolar accretion disk, outflows, and envelope including the processes operating
        in the early solar system together with the production of CAIs and dynamics of dust grains. Notice that the CAI factory is only
        operating at very early times, when the density in the envelope is still high, and there is mixing among the jet, inner wind,
        and envelope in the respective boundary layers. Timescales are for transport of dust grains.}
        \label{fig:sketch}
\end{figure}

A summary of the different processes and how they pertain to the circulation of matter in the early protosolar system\rev{, if CAIs are circulated by
outflows,} is shown Figure~\ref{fig:sketch}. CAIs are essentially unaltered since their formation. Therefore, they are ideal tracers of the dust dynamics in the early
solar system that can be compared directly to dust grains in our numerical models.

\rev{The existence of separate reservoirs recorded in the isotopic composition of meteorites, combined with the general gap-opening
properties of gas giants, has led several authors to suggest that Jupiter played a pivotal role in separating material in the early
solar system \citep{2011E&PSL.311...93W,VanKooten2011,Kruijer6712,2018E&PSL.498..257E,2018ApJS..238...11D}, but quantitative models
linking the cosmochemical constraints to physical disk models have not been explored \citep[except conceptually by][]{2018ApJS..238...11D}.
The dust gap opened by a giant planet is limited to larger particles, and the ability of the planet to filter dust depends on the
physical conditions in the disk and the size of the grains, encapsulated in the Stokes number introduced below. While passage
will be inhibited for large grains, small-enough grains stream through the gap as part of the gas flow.} Recently, \citet{2018ApJ...854..153W} presented
a detailed numerical study of the dust permeability of planet-induced gaps.
In the current paper, we specifically focus on the early solar system, extend the models to include two gas giants, and combine the
numerical models with cosmochemical evidence to probe the conditions near Jupiter in the protosolar accretion disk \rev{assuming that CAIs
trace the general flow of dust and gas from the envelope and outer protosolar disk.}

We have conducted a systematic search for CAIs in polished slabs extracted from the primitive L3.10 OC NWA 5697.
We made low-resolution scans with a $\mu$-X-ray-fluorescence scanner
to characterize the elemental distribution and identify any large CAIs over large surface areas. This is complemented by high-resolution
scans across smaller areas using a scanning electron microscope (SEM). Targeted measurements are used for
follow-up measurements on CAI candidates
found in the low-resolution scans, and four serendipitous high-resolution scans over different fields are conducted to search for smaller CAIs, resulting
in the detection of four new CAIs in an OC. Combining these newly identified inclusions with five CAIs in OCs already reported in the literature, 
this allows us to put refined limits on the size distribution and occurrence rate of CAIs in the inner solar system. Furthermore, the laboratory data
are combined with a suite of numerical disk models, which include gas, dust, and up to two gas giants --- Jupiter and Saturn ---
that inform us of the permeability of the gap(s) as a function of dust grain size and of important physical conditions in the disk.

With the combination of state-of-the-art numerical models and
systematic cosmochemical data, we can put limits on (i) the surface density and viscosity of the accretion disk where Jupiter formed, (ii) the
relative sizes of the dust reservoirs in the inner and outer solar system, respectively, and (iii) the initial orbital geometry of the gas giants
in the early solar system.

In section~\ref{sec:theory}, we list the governing equations of gas and dust dynamics, introduce the three planetary configurations that
are considered, and describe the applied numerical methods. Section~\ref{sec:data} introduces the laboratory techniques and analysis
methodology. In section~\ref{sec:results}, we present both numerical and laboratory results, and in section~\ref{sec:discussion} we discuss
possible implications for the early protosolar disk and the emergence of the gas giant planets, before concluding in section~\ref{sec:concl}.

\section{Protosolar disk model} \label{sec:theory}

As witnessed by observations of nearby protoplanetary disks \citep{2011ARA&A..49...67W}, the main constituents of the mass in a protoplanetary disk
are in gaseous form. We thus start by addressing the gas dynamics of the system.
More specifically, for the purpose of this study, we consider a height-integrated, two-dimensional protoplanetary disk model.

\subsection{Gas Dynamics}\label{subsec:gas}

The equations governing the dynamical evolution of the gas in the context of an enhanced viscosity framework are the continuity equation
and the Navier-Stokes equation:
\begin{eqnarray}
&&\frac{\partial\Sigma_\mathrm{g}}{\partial t} + \mathbf{\nabla}\cdot \left(\Sigma_\mathrm{g}\mathbf{u} \right)   \, = \, 0 \,, \label{eq:contgas}  \\[5pt]
&&\Sigma_\mathrm{g}\left( \frac{\partial  \mathbf{u}}{\partial t} + \mathbf{u}\cdot \mathbf{\nabla}  \mathbf{u}\right)
\, = \,   - \mathbf{\nabla}P - \mathbf{\nabla}\!\cdot\!\tau \; - \Sigma_\mathrm{g}\mathbf{\nabla \phi}  -  \Sigma_{\mathrm{d}} \mathbf{f}_{\mathrm{d}}\,.
\label{eq:NS-gas}
\end{eqnarray}
Here  $\mathbf{u}=(u_r,u_\varphi)$ denotes the gas velocity vector field, where $r$ and $\varphi$ represent coordinates along the
spherical radius and the azimuth, respectively. As appropriate for an efficiently radiating gas, the surface density, $\Sigma_\mathrm{g}$,
and the vertically integrated gas pressure, $P$, are assumed to be related via the isothermal sound speed, $c_\mathrm{s}$, according
to $P=\Sigma_\mathrm{g} c_\mathrm{s}^2$. Moreover, the viscous stress tensor has its standard form, that is,
\begin{equation}
\tau \equiv \Sigma_{\rm g}\, \nu \left[\; \mathbf{\nabla} \mathbf{u} + (\mathbf{\nabla}\mathbf{u})^T - \frac{2}{3}(\mathbf{\nabla}\cdot \mathbf{u})\,\mathds{1}\;\right]\,,
\end{equation}
where $\mathds{1}$ denotes the identity matrix and $\nu$ is the (enhanced) kinematic viscosity. The last term in the momentum equation is due to the frictional force $\mathbf{f}_{\mathrm{d}}$ exerted onto the gas by the presence of dust, itself described by a surface density $\Sigma_\mathrm{d}$.

We now consider the basic disk structure that results from assuming that the gravitational potential, $\phi$, is determined solely by the stellar mass $M_\ast$ at the coordinate origin, that is, $\phi = \phi_\ast \equiv -GM_\ast/r$, where $G$ is the gravitational constant.
To parameterize the level of disk turbulence in simple terms, we use the dimensionless  $\alpha$-parameter description of the viscosity \citep{1973A&A....24..337S}
\begin{equation}\label{eq:nu}
\nu \equiv \alpha c_{\rm s} h r\,.
\end{equation}
Here $h\equiv H_\mathrm{P}/r$ is the aspect ratio of the gas disk, which compares the pressure scale height $H_\mathrm{P}=c_\mathrm{s}/\Omega_\mathrm{K}$, where $\Omega_\mathrm{K}=\sqrt{GM_\ast/r^3}$ denotes the Keplerian angular frequency, to the orbital radius.
We further assume that  $\Sigma_\mathrm{d} \ll \Sigma_\mathrm{g}$  so that the so-called feedback term $-\Sigma_{\mathrm{d}} \mathbf{f}_{\mathrm{d}}$ in Equation~(\ref{eq:NS-gas}) is negligible\footnote{We note that even though the location of the outer pressure maximum created by a planetary gap changes with this approximation, the dust transport properties to the inner system remain mostly unaffected \citep{2018ApJ...854..153W}.}.

Under the previous considerations, we seek steady-state solutions of the equations of motion for the gas assuming a vertically isothermal disk and power-law profiles for both the surface density and temperature, such that
\begin{equation}\label{eq:powerlaw}
  c_\mathrm{s}^2(r) \propto r^{\bar{q}},\qquad
    \Sigma_{\mathrm{g}}(r) \propto r^{\bar{p}}\,.
\end{equation}
Following \citet{Pringle1981}, the gas surface density of a thin disk in steady state can be linked to a constant accretion rate, $\dot{M}$, and the viscosity, $\nu$, via
\begin{equation}\label{equ:gasdens}
  \Sigma_\mathrm{g} = \frac{\dot{M}}{3\pi \nu}\,.
\end{equation}
On the other hand, the accretion rate can also be written as the mass flux through a ring at a certain distance $r$ to the star as
\begin{equation}
  \dot{M} = - 2\pi r\, \Sigma_\mathrm{g}\, u_r\,,
\end{equation}
providing an expression for the radial gas velocity
\begin{equation}\label{equ:ur}
  u_r = -\frac{3\nu}{2r}\,.
\end{equation}
In the case of Keplerian rotation, we require $\bar{q} + \bar{p} + 3/2 = 0$ for the accretion rate to be constant as a function of radius.
In general, the azimuthal velocity of the gas deviates from the Keplerian angular velocity depending on the pressure structure of the disk. The steady-state solution of Equation~(\ref{eq:NS-gas}) when neglecting the contribution of the viscous term is
\begin{equation} \label{equ:gasvphi}
  u_\varphi = v_\mathrm{K} \sqrt{1-\eta}\,,
\end{equation}
with $\eta \equiv -h^2(\partial \log{P}/\partial \log{r}) \sim h^2$ a small parameter describing the deviation from the Keplerian velocity $v_\mathrm{K} = r \,\Omega_\mathrm{K}$.

\subsection{Dust Dynamics}\label{subsec:dust}

We treat dust as a pressureless fluid, which enables us to describe its motion by the Euler equations:
\begin{eqnarray}
&&\frac{\partial\Sigma_\mathrm{d}}{\partial t} + \mathbf{\nabla}\cdot \left(\Sigma_\mathrm{d}\mathbf{v} +\mathbf{j}\right)  =  0 \,, \label{eq:contdust}\\[4pt]
&&\Sigma_\mathrm{d} \left( \frac{\partial  \mathbf{v}}{\partial t} + \mathbf{v}\cdot \mathbf{\nabla}  \mathbf{v}\right)\;  =  \; -\Sigma_\mathrm{d}\mathbf{\nabla \phi} + \Sigma_{\mathrm{d} } \mathbf{f}_{\mathrm{d}}\,.
\label{eq:NS-dust}
\end{eqnarray}
Here $\mathbf{v}=(v_r,v_\varphi)$ is the dust velocity vector field and $\mathbf{j}$ represents a diffusive flux driven by the gradient in the dust concentration \citep{1984ApJ...287..371M}:
\begin{equation}
\mathbf{j} = -D_{\rm d}\Sigma\, \mathbf{\nabla} \left( \frac{\Sigma_\mathrm{d}}{\Sigma}\right)\,,
\label{eq:diffflux}
\end{equation}
with the combined surface density $\Sigma =  \Sigma_\mathrm{g} + \Sigma_\mathrm{d}$.
The diffusion coefficient $D_{\rm d}$ defines the level of diffusivity and is assumed to be equal to the viscosity of the gas $D_{\rm d}= \nu$, which is a sensible approximation for small grains \citep{2007Icar..192..588Y}.

The equations of motion for the dust differ from those describing the gas, which can lead to a discrepancy in their velocities and consequently to friction. In general, the friction force exerted by the gas onto the dust constitutes a significant contribution to the dust dynamics, also in cases where it is negligible for the gas. For dust grains whose size, $a$, is small enough compared to the mean free path between gas molecules, $\lambda_\mathrm{mfp}$, this interaction can be regarded as independent collisions for which the coupling of dust and gas is described by the Epstein drag law. This description is valid if $\lambda_\mathrm{mfp}/a \gtrsim 4/9$ \citep{1977MNRAS.180...57W}, which is fulfilled in our case ($\lambda_\mathrm{mfp} = 1/(n \sigma) \approx 100$ cm). Here  the frictional force depends linearly on the difference in velocities $ \fdrag \equiv C_{\rm drag} (\vvg - \vvd)$ (\citeauthor{Safronov1972} \citeyear{Safronov1972}, p.~15; \citeauthor{1972fpp..conf..211W} \citeyear{1972fpp..conf..211W}), and on the drag coefficient $C_\mathrm{drag}=\Omega_\mathrm{K}/\St$, which determines the coupling strength. $C_\mathrm{drag}$ depends on the Stokes number. As we consider a height-integrated two-dimensional disk, we use the Stokes number evaluated at the midplane
\begin{equation}\label{equ:Stokes}
\St = \frac{a\rho_\mathrm{mat}}{\Sigma_\mathrm{g}}\frac{\pi}{2}\,,
\end{equation}
where $\rho_\mathrm{mat}$ is the material density of the solids. When $\St \ll 1$ there is a large contribution of the drag force, hence coupling the motion of the dust tightly to that of the gas, whereas when $\St \gg 1$ the drag force becomes negligible, and the dust decouples from the gas.

The velocity field for dust in steady-state motion within the gaseous disk structure derived above can be approximated by \citep{2002ApJ...581.1344T}
\begin{eqnarray}
\displaystyle{ v_{r}} & \approx & \displaystyle{\frac{\St^{-1} u_{r} -\eta \, v_\mathrm{K}}{\St+\St^{-1}}}\,,
\label{eq:dustvr}\\[4pt]
\displaystyle{ v_{\varphi}} & \approx & \displaystyle{u_\varphi - \frac{1}{2}\St\, v_r}\,,
\label{eq:dustvphi}
\end{eqnarray}
This shows explicitly that the dust motion depends strongly not only on the gas dynamics (via the coupling to $u_r$ and $u_\varphi$) but also
on the gas density distribution (via the dependence of $\St$ on $\Sigma_\mathrm{g}$).

\subsection{Planet--Disk Interaction}\label{subsec:pd}
Up to this point, we have discussed the disk structure that results from considering only the central stellar gravitational potential, which we refer to as the steady-state unperturbed gas disk. In the presence of one or more planets in the disk, the gravitational potential in Equations~(\ref{eq:NS-gas}) and (\ref{eq:NS-dust}) becomes
\begin{equation}
\phi = \phi_\ast + \sum_i \phi_{\mathrm{p}, i}\,,
\end{equation}
where $\phi_{\mathrm{p}, i}$ is the potential due to planet $i$.
In our treatment we consider the planets to be point sources of mass  $m_\mathrm{p}$ rotating in a Keplerian fashion at the fixed radius $\mathbf{r}_\mathrm{p}$, that interact with the disk solely through their gravitational potential
\begin{equation}\label{equ:planet}
\phi_{\mathrm{p}, i} = -\frac{G\,m_{\mathrm{p}, i}}{\left(|\mathbf{r}-\mathbf{r}_{\mathrm{p}, i}|^2+\epsilon^2\right)^{\frac{1}{2}}} + \frac{G\,m_{\mathrm{p}, i}}{r_{\mathrm{p}, i}^2}r\cos{\varphi}\,.
\end{equation}
Here the parameter $\epsilon$ is the smoothing length, which takes into account the (integrated) vertical extent of the disk for our two-dimensional model \citep{Masset2002,Mueller2012} and avoids the singularity at $\mathbf{r}=\mathbf{r}_\mathrm{p}$. The second term in Equation~(\ref{equ:planet}) is the so-called indirect term and corresponds to a noninertial acceleration term due to the astrocentric frame of reference used.

\subsection{Models of consideration}\label{subsec:setups}
We are interested in the inward transport of solid material through the orbital location of Jupiter in the early solar system. Therefore, we set up the gas disk accordingly. Our fiducial model corresponds to a nonflared disk ($h=\mathrm{const}$), for which we list the set of parameters in Table~\ref{tab:para}.
\begin{table}
        \centering
        \caption{\textrm{Fiducial model parameters.}}
        \label{tab:para}
        \begin{tabular}{l c r }
                \hline\hline
                Viscosity parameter \qquad  \qquad & $\alpha$   \qquad& $3\times10^{-3}$ \\
                Surface density slope \qquad  \qquad &$\bar{p}$   \qquad& -0.5 \\
                Temperature slope \qquad  \qquad & $\bar{q}$   \qquad& -1.0 \\
                Aspect ratio \qquad  \qquad&$h$   \qquad& 0.05 \\
                Stellar mass \qquad  \qquad &$M_\ast\,[\mathrm{M}_\odot]$  \qquad & $1$ \\
                Accretion rate \qquad  \qquad \quad& $\dot{M}\,[ \mathrm{M}_\odot \mathrm{yr}^{-1}]$ \quad \qquad & $10^{-7}$ \\
                Material density \qquad  \qquad \quad& $\rho_\mathrm{mat}\,[ \mathrm{g} \,\mathrm{cm}^{-3}]$ \quad \qquad & $3$ \\
                Smoothing length \qquad  \qquad \quad& $\epsilon\,[ H_\mathrm{p}]$ \quad \qquad & $0.6$ \\

                \hline
        \end{tabular}
\end{table}

In this environment we investigate the effects of three different planetary configurations. Setup I includes the gravitational potential of Jupiter in the disk located at the same distance to the star as it is today. Setup II adds Saturn to the configuration, also at its current orbital distance. \rev{To highlight the effect of a different radial spacing between Jupiter and Saturn, in Setup III we consider them to be in a 3:2 mean motion resonance (MMR) to study the impact of this scenario on the radial transport of solids.} For this we reduce the orbital distance of Saturn accordingly. Setups II and III are motivated not only to see the effect of Saturn but also to compare different models of planet migration. Setup II implicitly assumes that Saturn formed \textit{in situ} or at least ended up at its current location very early on; Setup III, on the other hand, is motivated by the so-called Nice II model, where Saturn migrated radially inward until it got locked in a resonance with Jupiter \citep{2011AJ....142..152L}.

\subsection{Numerical setup}
We solve the disk equations with embedded planets using the publicly available code FARGO3D\footnote{\href{http://fargo.in2p3.fr}{http://fargo.in2p3.fr}} \citep{Benitez-Llambay2016} in its multifluid extension accounting for dust-to-gas coupling according to \citet{2018arXiv181107925B}. We treat the multiple dust species as pressureless fluids. This code solves the Navier-Stokes equations using finite-difference upwind, dimensionally split methods, combined with the FARGO algorithm \citep{Masset2000} for the orbital advection and a fifth-order Runge-Kutta integrator for planetary
orbits. The numerical setup used in this paper is similar to that presented in \citet{2018ApJ...854..153W}.
We solve the equations describing our problem on a two-dimensional polar grid divided in 1024 uniform azimuthal sectors. The radial domain of the disk is set in the range $r/r_{\rm Jup}\in[0.25,3.00]$, uniformly divided over 512 cells. To be able to evolve a large set of models over ten thousands of orbits, we ran the code exclusively on GPUs using the University of Copenhagen HPC facility.

Because we are interested in finding out how much dust of a putative reservoir that was initially located in the outer solar system can be transported toward the inner solar system, we initialize the dust distribution only \textit{outside} of the gaps carved by the planets:
\begin{equation}
\Sigma_{\rm d}(r) = \left\{\begin{array}{ll}
10^{-20}  & \qquad\mbox{if } r<\tilde{r} \\
\varepsilon\,\Sigma_\textrm{g} & \qquad\mbox{if } r\geq \tilde{r}
\end{array}\right.\,.
\label{eq:dustinit}
\end{equation}
Here $\varepsilon$ is the dust-to-gas ratio that is set to $\varepsilon=0.01$. Because the influence of dust onto the gas is neglected, this value does not qualitatively influence the outcome. In Equation~(\ref{eq:dustinit}) the position $\tilde{r}$ is chosen such that it is well outside the gap of the corresponding setup. For Setup I we hence chose $\tilde{r} = 1.5\times r_\mathrm{Jup}$, and for Setups II and III we chose $\tilde{r} = 2.5 \times r_\mathrm{Jup}$, where $r_\mathrm{Jup}$ is Jupiter's semi-major axis.

We implement boundaries for dust and gas separately because their mass fluxes can be vastly different. We set the outer boundaries of the disk equal to the equilibrium disk solutions for densities and velocities (see sections~\ref{subsec:gas} and \ref{subsec:dust}). We do the same at the inner boundary for the velocities but adopt a zero-gradient boundary for gas and dust densities. In order to avoid spurious reflections of the spiral wake generated by the planets, we adopt damping buffer zones according to \citet{Val-Borro2006}.
\newline

We study the evolution of the dust for different disk viscosities and the three different setups that are summarized in Table~\ref{tab:setups}. We varied the number of Jupiter orbits for every setup and viscosity level in order to analyze the dust density profiles once a steady state has been reached in the inner disk. The required number of Jupiter orbits depends on the initial location of the dust reservoir $\tilde{r}$ and the viscous evolution timescale of the disk.
The efficient use of GPUs enabled us to run the simulations for the required many thousands of orbits while including several dust species at once.

\begin{table}
        \centering
        \caption{\textrm{Summary of simulation sets.}}
        \label{tab:setups}
        \begin{tabular}{lccc}
                \hline\hline
                Scenario$\alpha$ &  Orbits & Setup \\ \hline
                Jupiter & $3\times10^{-4}$ & 45,000& \\
                Jupiter & $6\times10^{-4}$ & 20,000 &\\
                Jupiter & $10^{-3}$ & 20,000& I \\
                Jupiter & $3\times10^{-3}$ & 10,000 &\\
                Jupiter & $10^{-2}$ & 10,000& \\ \hline
                J + S ~~(in situ) & $3\times10^{-4}$ & 45,000& \\
                J + S ~~(in situ) & $6\times10^{-4}$ & 20,000 &\\
                J + S ~~(in situ) & $10^{-3}$ & 20,000 &II\\
                J + S ~~(in situ) & $3\times10^{-3}$ & 20,000& \\
                J + S ~~(in situ) & $10^{-2}$ & 10,000 &\\ \hline
                J + S ~~(in resonance) & $3\times10^{-4}$ & 45,000 &\\
                J + S ~~(in resonance) & $6\times10^{-4}$ & 20,000 &\\
                J + S ~~(in resonance) & $10^{-3}$ & 20,000 &III\\
                J + S ~~(in resonance) & $3\times10^{-3}$ & 20,000& \\
                J + S ~~(in resonance) & $10^{-2}$ & 10,000 &\\
                \hline
        \end{tabular}
\end{table}

\section{Cosmochemical evidence}\label{sec:data}

Primitive meteorites contain a number of components: nanometer- to micrometer-sized fine-grained partly presolar material, termed the matrix;
micrometer- to millimeter-sized molten spherical chondrules formed by transient heating events in the protosolar nebular; and
high-temperature condensates, in particular CAIs, which originally formed as condensates from a hot metal rich gas
at the birth of the solar system. The inventory and distribution of material in the early solar system are shaped by the physical processes active in the
protosolar disk, the transport of material in the protosolar system, and the raw material supplied in the accretionary flow.

To constrain and compare our numerical models with the laboratory data, we aim to identify material that has been transported from
the outer to the inner solar system.
Chondrules originating from different reservoirs in the disk can be identified through isotopic measurements that require a significant
laboratory effort. The identification of different components in the chondrites is comparatively much simpler and can be done in situ using X-ray fluorescence and SEM measurements to generate a combination of 
elementary maps and backscattering images. As such, we have used these techniques to conduct a systematic search for CAIs in slabs from the 
OC NWA 5697. 

\subsection{Laboratory techniques and data acquisition}
\begin{figure}
        \centering
        \includegraphics[width=\columnwidth]{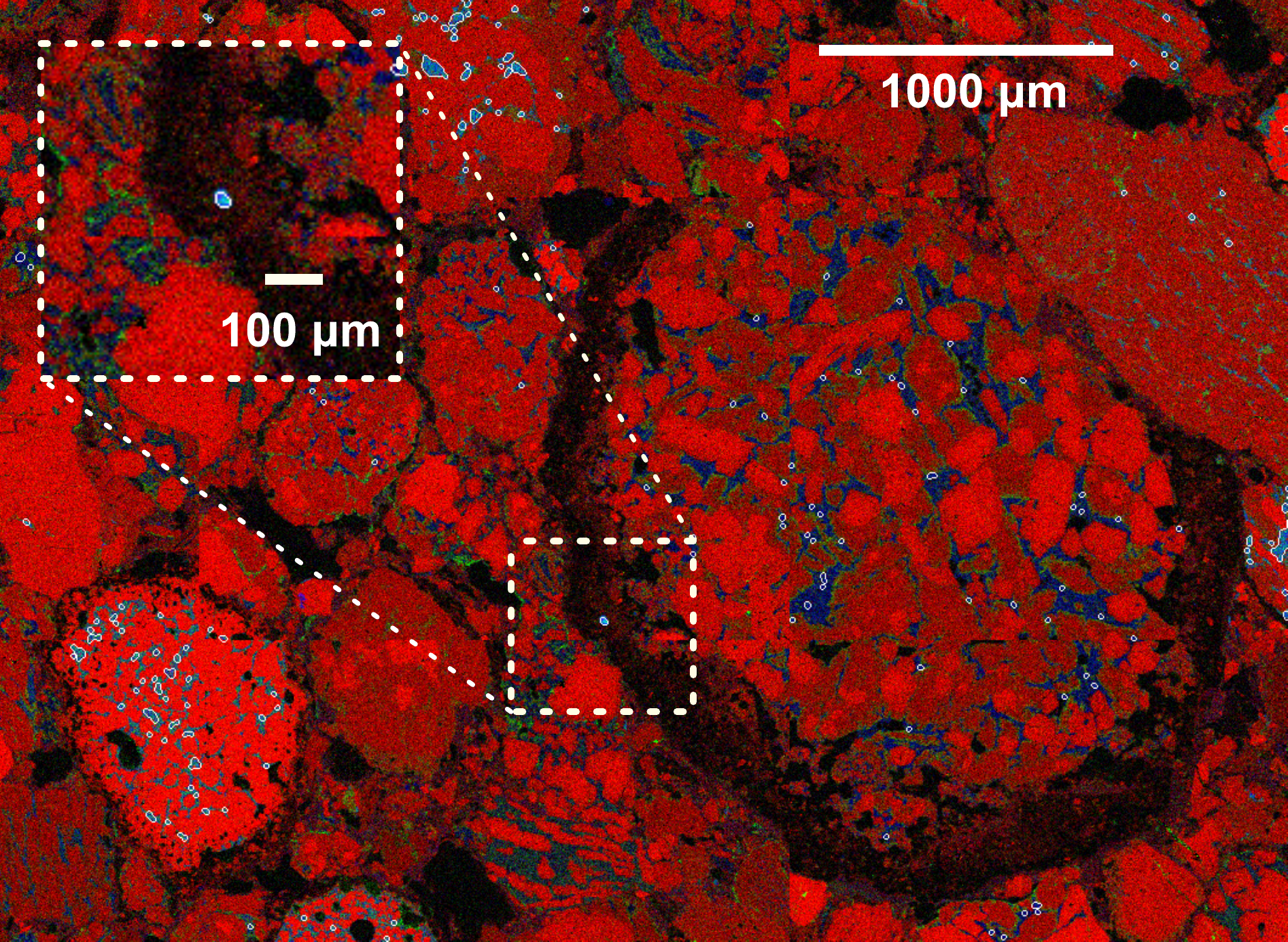}
        \caption{Illustration of the automatized selection procedure. Magnesium (red), calcium (green), and aluminum (blue) SEM scan in
        the area surrounding NWA 5697-3. Pixel size is 4.13 $\mu$m.
        The white lines mark areas with aluminum-rich material. At the center of the marked box is a small CAI hiding in an accretionary
        rim of a larger chondrule. See Figure \ref{fig:scans} for a further zoom-in of NWA 5697-3.}
        \label{fig:small-cai}
\end{figure}

We used a Bruker Tornado $\mu$-X-Ray-fluorescence ($\mu$-XRF) scanner outfitted with a 30 mm$^2$ silicon drift detector energy
dispersive spectrometer (EDS) and a Philips XL 40 SEM outfitted with a Thermo Nanotrace 30 mm$^2$
lithium doped silicon EDS to perform X-ray emission mapping on two different scales. 

The Tornado $\mu$-XRF is hosted locally at the
Quadlab research group at the Natural History Museum of Denmark. It provides comparatively rapid and high-sensitivity mapping by
scanning the sample stage under an intense primary X-ray beam, generated by a 50 keV 230 $\mu$A Rh tube. The X-rays are focused
by polycapillary optics to a $\sim$25 $\mu$m spot at the focal point. This generates X-ray fluorescence within the sample, typically
providing 70,000--150,000 cps in the detector while scanning broadly chondritic materials. We used a pixel dwell time of $\sim$50--100 ms,
corresponding to typically $\sim$6000 counts pixel$^{-1}$.

The XL 40 SEM at the Geological Survey of Denmark and Greenland electronically
scans a focused $\le$1 $\mu$m 17 KeV and $\sim$60 $\mu$A electron beam across the sample surface, generating X-ray emission from a
comparatively smaller region of the sample. This typically generates $\sim$10,000-20,000 cps in the EDS, while simultaneously providing
useful high spatial resolution backscatter electron (BSE) information from the surface. 

For the $\mu$-XRF, the X-ray interaction volume and
hence the inherent spatial resolution are larger than the respective primary beams, due to, e.g.,~scattering and fluorescence effects within the materials.
The effective spatial resolution is largely dependent on the characteristic X-ray energy being monitored, following their absorption coefficients
within the sample matrix, a parameter that broadly increases at lower energies. This implies that typical $k-\alpha$ emission for light elements
is emitted from a smaller volume than equivalent $k-\alpha$ emission from heavier elements. We estimate that the effective resolution in the
$\mu$-XRF and the SEM is approximately 30 and 2 $\mu$m, respectively, for Mg, Al, and Si, and comparatively higher for Ti and Fe.
The lower resolution for Fe has implications for our CAI search algorithms in the SEM maps because the stray Fe $k-\alpha$ X-rays from neighboring
materials such as Fe-rich matrix are more prone to appear during analysis of essentially Fe-free materials.

\subsection{CAI screening procedure and data analysis}

Given the closed proprietary file format and the limited automatic quantification capabilities of the Bruker $\mu$-XRF software, we chose to visually inspect the maps for Al-, Ca- and Ti-rich materials. Since this procedure was not sufficient to confidently identify actual CAIs, all candidate materials were subsequently mapped using higher-resolution SEM EDS for final determination. We scanned and checked candidate materials from a total of  $\sim$53 cm$^2$ scanned with the $\mu$-XRF across three meteorite slabs. 

The spatial resolution of the $\mu$-XRF was furthermore insufficient to identify small CAI candidates, and we therefore also carried out four medium-resolution $\sim$5$\mu$m pixel$^{-1}$ SEM EDS mappings totaling 4.1 cm$^2$ in order to identify such materials, and we reimaged candidates at higher resolution for follow-up and confirmation. Given the large resulting data size and the goal of making a complete inventory even at the smallest CAI sizes, we supplemented manual candidate identification with an automated screening algorithm.

We exported the data in the form of elemental weight abundance maps, assuming typical oxide stoichiometry, i.e., Al in the form of Al$_2$O$_3$, Ca in the form of CaO, and Mg in the form of MgO. The subsequent automated analysis of the obtained elemental maps was done in two steps. In the first step we identified automatically pixels that could contain
refractory material according to cuts that characterize the elemental properties of CAIs. The following conditions are very generous in the
selection of candidate refractory material  and therefore guaranteed that we include all possible candidates:

\begin{itemize}
\item{The mass fraction of Al has to be above 10\%, and the relative mass fraction of Al to Mg has to be larger than 1.5. This selects aluminum rich material}
\item{The mass fraction of Fe has to be below 1\%. This is to find material that formed under reducing conditions.}
\end{itemize}
Once all candidate pixels have been selected, we remove all isolated pixels that have less than two neighboring selected pixels
and expand the resulting areas with two layers and additionally several more layers to make areas convex and reduce the noise. This results
in areas that are at least four pixels across, corresponding to a minimum length scale of 20 $\mu$m.
The marked pixels are then inspected by hand to find candidates for refractory inclusions based on
the texture and context of the material. The large majority of the selections are metastasis in chondrules and aluminum-rich chondrules, which can readily
be excluded. This enables us to reliably identify candidate inclusions with a diameter down to 20 $\mu$m or 4 pixels in the electron microscopy images.
The candidates are then reexamined in targeted follow-up SEM images with longer integration times and an elemental map resolution of less than
1 $\mu$m.
An example of an SEM image with refractory material selected with white contours is shown in Figure \ref{fig:small-cai} highlighting
the automatic marking of a small CAI that is hiding in the accretionary rim of a larger chondrule.

\section{Results}\label{sec:results}
\subsection{Numerical models}
\begin{figure}
        \centering
        \includegraphics[width=\columnwidth]{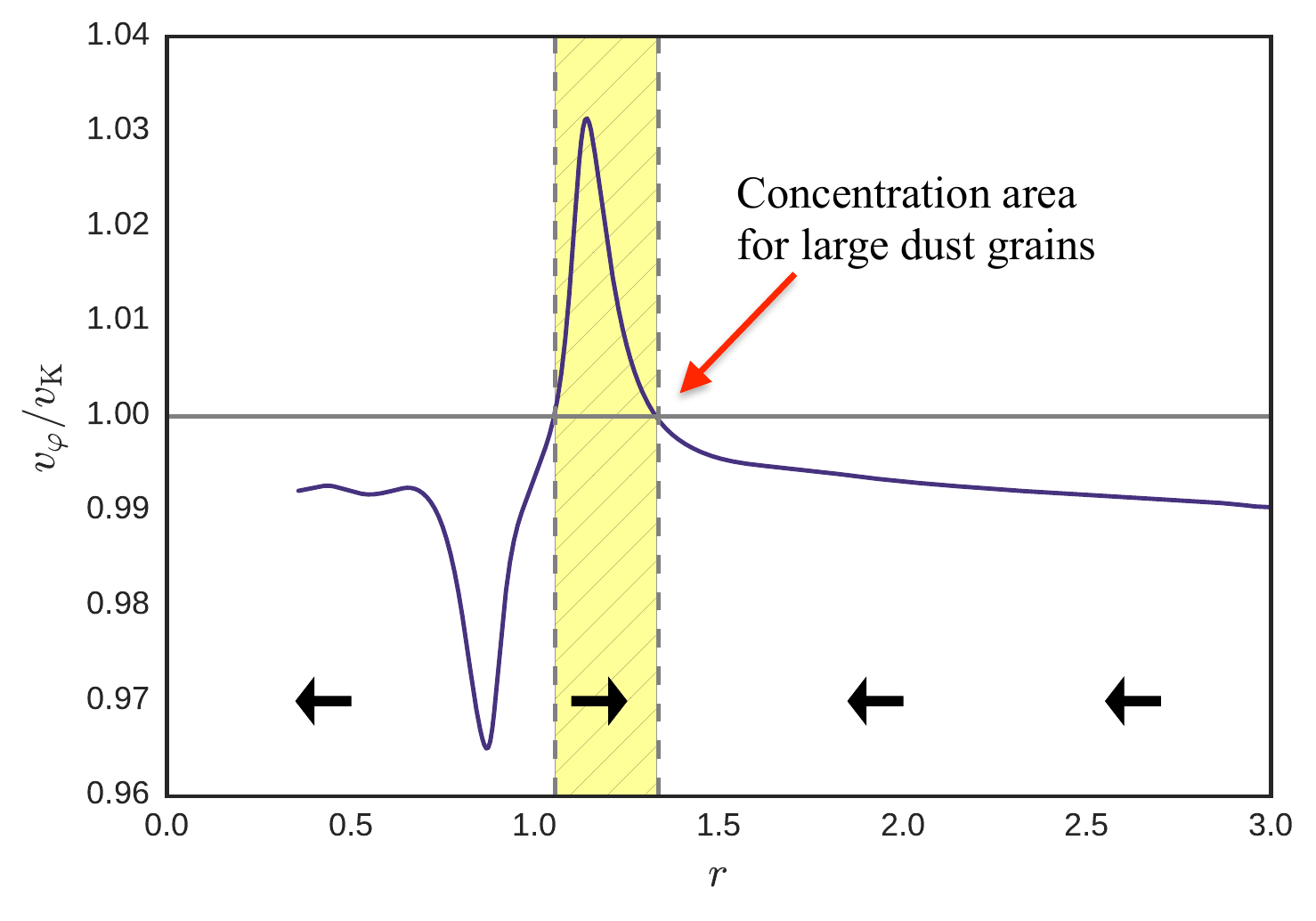}
        \caption{Azimuthal velocity profile of the gas disk in its azimuthal average. To highlight sub- and super-Keplerian regions in the disk, the velocity is illustrated in comparison to the Keplerian velocity, and arrows are added to illustrate the effect on the dust momentum. The snapshot is taken from our fiducial model simulation of Jupiter after 10,000 orbits. The planet is located at $r=1.0$.}
        \label{fig:azvel}
\end{figure}

A giant planet embedded in a protoplanetary disk exerts a gravitational torque onto its surrounding environment. The presence of the planet perturbs the azimuthal velocity $u_\varphi$ of the gas from its sub-Keplerian power-law profile from Equation~(\ref{equ:gasvphi}). Figure~\ref{fig:azvel} shows that just outside of the planet's location there is a sector in which the gas velocity becomes super-Keplerian. While gas that is orbiting with a sub-Keplerian velocity decelerates the slightly decoupled dust and makes it spiral inward  toward  the star \citep{1976PThPh..56.1756A,1977MNRAS.180...57W} a super-Keplerian angular velocity of the gas produces the opposite effect. Here  the friction exerted from the gas onto the dust adds a positive contribution in the radial and angular components of the dust velocity, seen in Equations (\ref{eq:dustvr}) and (\ref{eq:dustvphi}). If this contribution becomes large enough in comparison to the drag of the in-spiraling accreting gas, the radial drift of the dust grains will stop at this distance to the Sun, and if this condition is persistently complied with, one speaks of the dust being trapped. Whether a dust trap is established depends on the strength of the dust diffusion  the Stokes number of a dust grain, and therefore on its size and on the underlying gas disk structure.
To ensure that the planetary gap in the gas structure is in a quasi-steady state, we first run each simulation for 10,000 orbits before we introduce
the dust into the disk.
We explore the consequences for dust transport in the three different scenarios described in section \ref{subsec:setups}.
\begin{figure}[!th]
        \centering
        \includegraphics[width=\columnwidth]{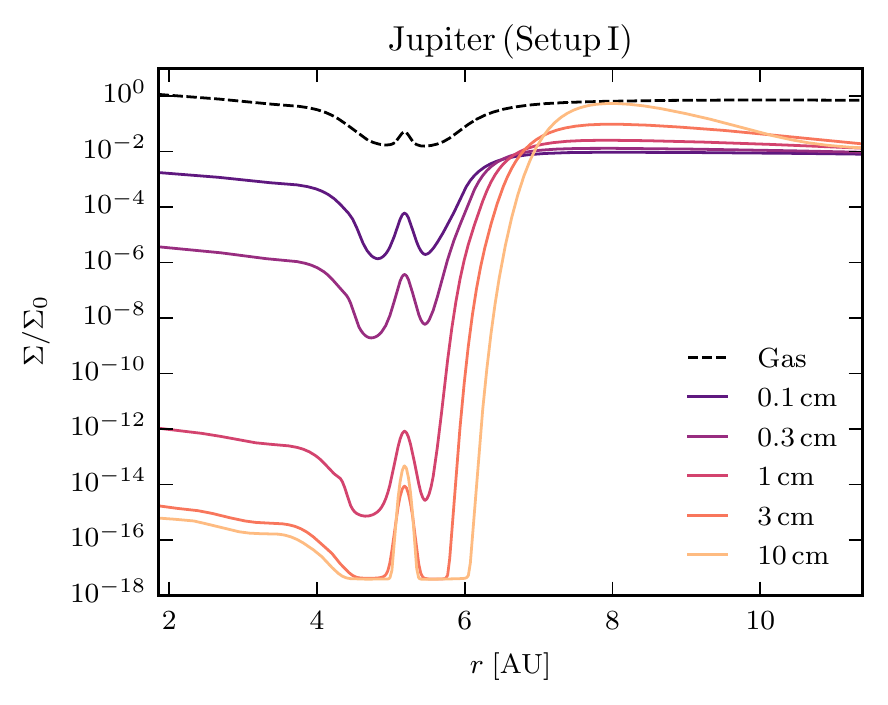}
        \caption{Gas and dust surface density profiles after 10,000 Jupiter orbits for $\alpha = 3\times10^{-3}$. The gas and dust surface densities
        are normalized to $\Sigma_0$, the gas surface density at Jupiter's location in the unperturbed state of the disk. Notice that the numerical floor at
        a relative density of $\approx 10^{-18}$ does not have any consequences for the results.}
        \label{fig:diffnodiff}
\end{figure}

\subsubsection{Jupiter acting alone}
The first case considers the effect of Jupiter on the transport of dust. After evolving the gas disk including Jupiter for 10,000 orbits at the planet location, we introduce the dust following Equation~(\ref{eq:dustinit}).
Figure~\ref{fig:diffnodiff} illustrates the size-dependent surface density structure of the dust after a further 10,000 orbits of dynamical dust and gas evolution. Note that the surface densities are normalized to the gas surface density at Jupiter's location in the unperturbed state of the disk, $\Sigma_0 = \Sigma_\mathrm{g}(r_\mathrm{Jup},t=0)$. The plot displays the size dependency of the transport to the inner solar system. For illustration we choose a reasonable size range of dust grains, to sample the regime where the dust grains start to become trapped at the gap edge.
The plot shows that small dust species almost retain the initial dust-to-gas ratio in the inner disk, while for larger sizes there is a clear depletion in the inner parts.

To quantify this effect, \citet{2018ApJ...854..153W} introduced a depletion factor, $\zeta$, as a function of dust size $a$. This factor $\zeta$ describes the ratio between the dust surface density at a location in the inner system (which in this study is set to $r_\mathrm{in} = 1.85 \mathrm{AU}$) after filtration by Jupiter and the dust surface density, $\hat{\Sigma}_\mathrm{d}$, of a disk without a planet:
\begin{equation}
        \zeta(a) = \left. \frac{\Sigma_\mathrm{d}(a)}{\hat{\Sigma}_\mathrm{d}(a)}\right|_{r_{\rm in}}\, .
        \label{eq:depletion}
\end{equation}
This allows the quantified comparison of several cases considering different viscosities. Note that, since we keep the stellar accretion rate fixed throughout the simulations, Equation~(\ref{equ:gasdens}) implies that $\Sigma_0 \propto \alpha^{-1}$, which gives us a direct connection between the gas surface density and the viscosity.

\begin{figure}
        \centering
        \includegraphics[width=\columnwidth]{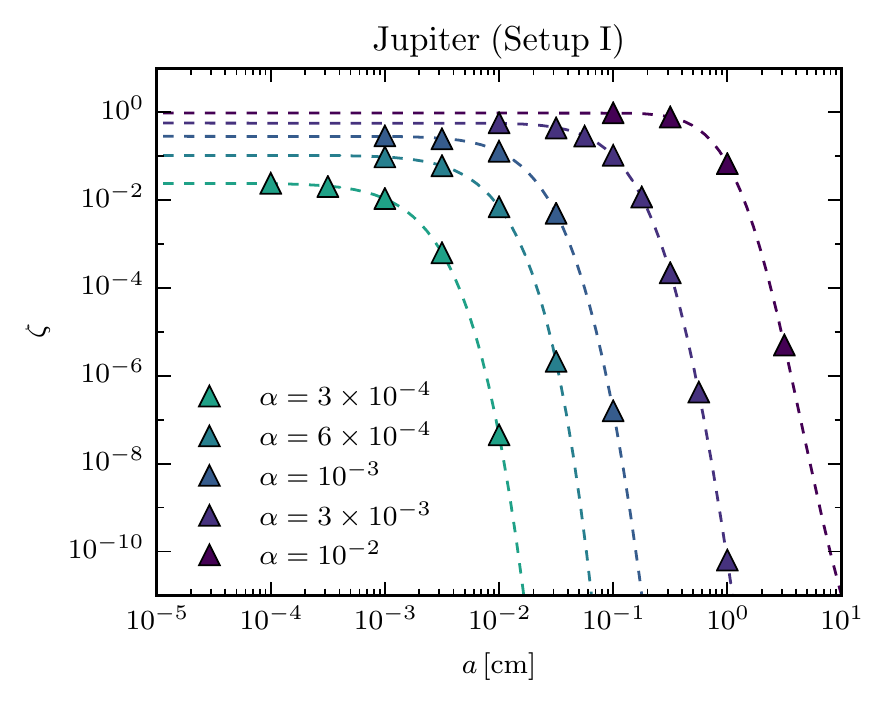}
         \caption{Viscosity dependence of filtration. The figure shows the depletion factor for filtration by Jupiter (Setup I). The number of planetary orbits of each simulation can be looked up in Table~\ref{tab:setups}. The dashed lines show the corresponding fit to the data points, for which we assumed an error function (see Equation~\ref{eq:fit}). The parameters of this fit can be taken from Table ~\ref{tab:fit}.}
        \label{fig:filter}
\end{figure}

\begin{table}
        \centering
        \caption{\textrm{Parameters for Equation~(\ref{eq:fit}).}}
        \label{tab:fit}
        \begin{tabular}{lccccc}
                \hline\hline
                Setup&$\alpha$&$A$&$B$&$C$ &  $D$  \\ \hline
                I&$3\times10^{-4}$&$6.9\times10^4$  &-0.54 &  4.16 & $-6.9\times10^4 $\\
                \quad&$6\times10^{-4}$&$4.3\times10^5$ & -0.48  & 5.60&  $-4.3\times10^5$\\
                \quad&$10^{-3}$&$11.5$ & -1.13  & -0.15  &$-12.0$\\
                \quad&$3\times10^{-3}$&9.5  &-1.25  &0.15 &-9.7 \\
                \quad&$10^{-2}$& 15.8 &-1.10 &1.18&-15.9 \\
                \hline
                II&$3\times10^{-4}$&$7.3\times10^4$  &-0.62 &  2.07 & $-7.3\times10^4 $\\
                \quad&$6\times10^{-4}$&$1.1\times10^6$ & -0.65  & 2.97&  $-1.1\times10^6$\\
                \quad&$10^{-3}$&$1.3\times10^5$ & -0.57  & 3.49  &$-1.3\times10^5$\\
                \quad&$3\times10^{-3}$&9.9  &-1.96  &-0.92 &-10.2 \\
                \quad&$10^{-2}$& 9.9 &-2.42 &-0.49 &-10.0 \\
                \hline
                III&$3\times10^{-4}$&$8.8$  &-1.18& 0.04 & $-9.3$\\
                \quad&$6\times10^{-4}$&$3.7\times10^5$ & -0.69  & 4.25&  $-3.7\times10^5$\\
                \quad&$10^{-3}$&$13.4$ &-1.19  & 0.69 &$-13.3$\\
                \quad&$3\times10^{-3}$&8.6  &-1.53  &0.76 &-8.3 \\
                \quad&$10^{-2}$& 0.7 &-2.35 &0.98&-0.5 \\
                \hline

        \end{tabular}
\end{table}

\begin{figure*}
        \centering
        \includegraphics[width=\textwidth]{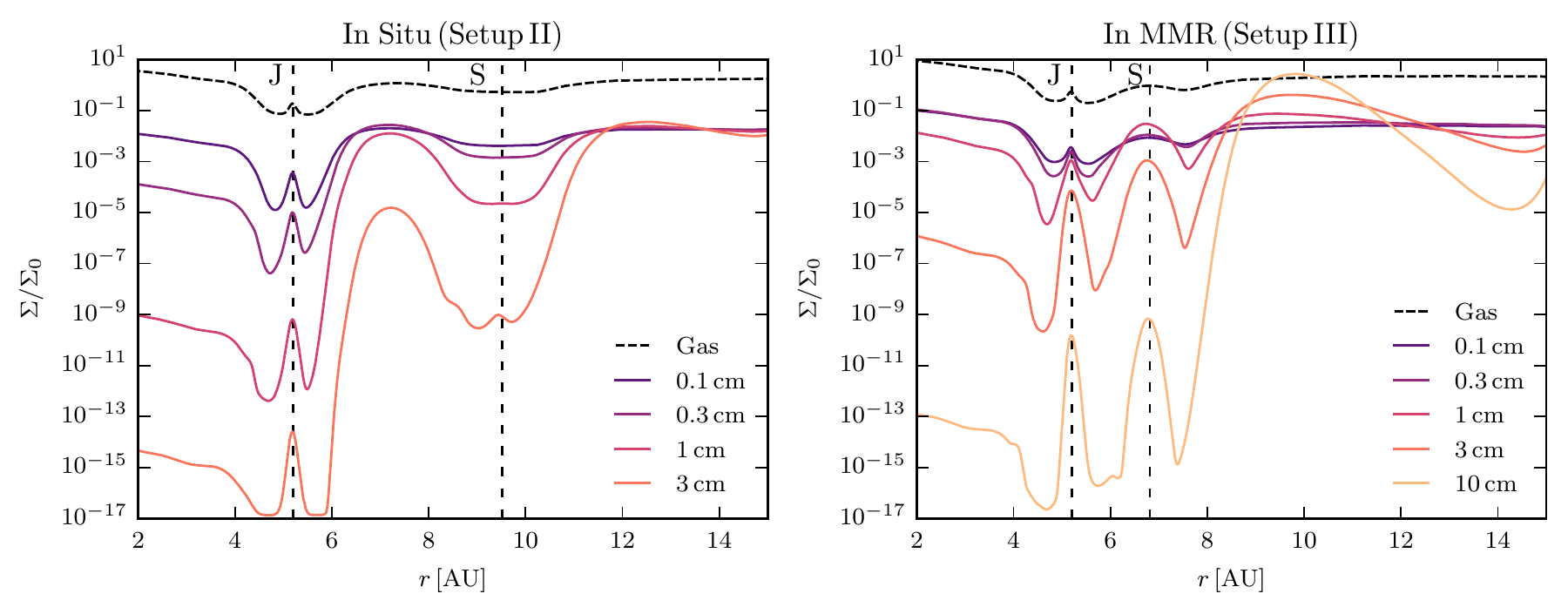}
        \caption{Gas and dust surface density profiles after 10,000 Jupiter orbits for $\alpha = 3\times10^{-3}$. The left panel shows the outcome with Jupiter and Saturn at the radial distances to the Sun of today, whereas the right panel shows them in a 3:2 orbital resonance. Note  that the range of grain sizes changes one order of magnitude between the two plots \rev{(notice that because 3 cm-sized particles are already effectively filtered out in Setup II, the larger grain size was omitted)}.}
        \label{fig:js-jsr}
\end{figure*}

\begin{figure*}
        \centering
        \includegraphics[width=\textwidth]{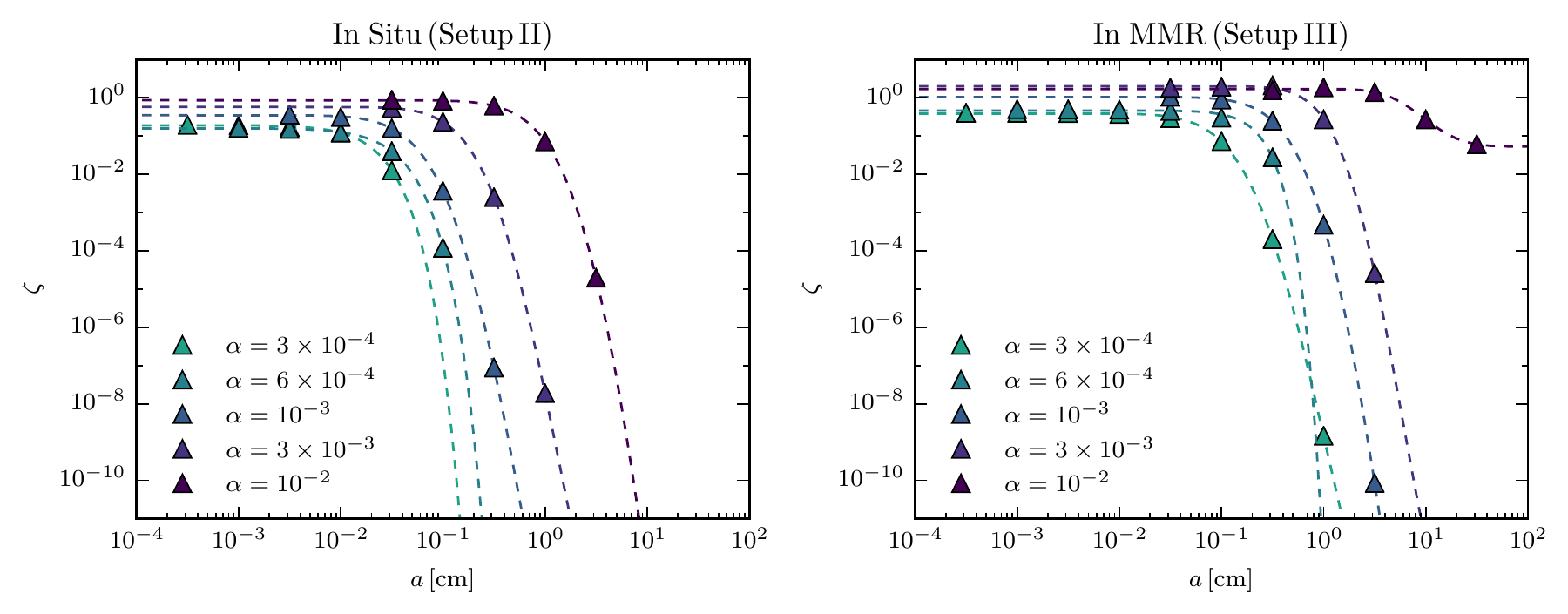}
        \caption{Viscosity dependence of filtration. The figure shows the depletion factor for filtration by Jupiter and Saturn (Setups II and III).
        The number of planetary orbits of each simulation can be looked up in Table~\ref{tab:setups}. The dashed lines show the corresponding fit to the
        data points, for which we assumed an error function (see Equation~\ref{eq:fit}). The parameters of this fit can be taken from Table ~\ref{tab:fit}.
        }
        \label{fig:js-filter}
\end{figure*}
\begin{figure}
        \includegraphics[width=\columnwidth]{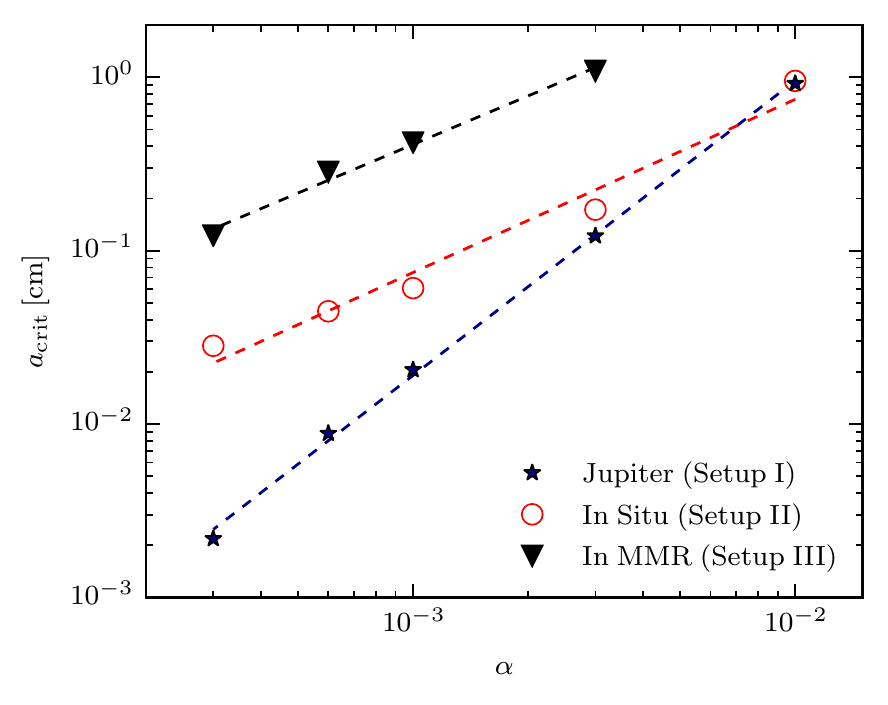}
        \caption{The $\alpha$-dependence of the critical grain size above which filtration becomes efficient. The plot directly compares the three different setups. Dashed lines show a best-fit power laws according to the corresponding formula of Equations~(\ref{equ:acrit1})--(\ref{equ:acrit3}). In the case of MMR of Jupiter and Saturn, no critical grain size was obtained for $\alpha=10^{-2}$.}
        \label{fig:js-alpha}
\end{figure}

\begin{table*}[t]
\centering
\caption[]{CAIs in ordinary chondrites}
\label{tab:caifrags}
\begin{tabular}{lllll}
\noalign{\smallskip}
\hline
\hline
\noalign{\smallskip}
Chondrite & Areal Size [$\mu$m$^2$] & Type & Reference & Comments \\
\noalign{\smallskip}
\hline
\noalign{\smallskip}
NWA 8276-1 & 600 $\times$ 400 & & \citet{2016LPI....47.1989R} & Irregular \\
Semarkona 1805-2-1 &$\sim$480 max dim &  & \citet{1984GeCoA..48..693B}  & Fragmented rim, lens shaped \\
Semarkona 4128-3-1 & 320 $\times$ 130 & Type A & \text{\citet{2001M&PS...36..975H}} &  Irregularly shaped \\
Moorabie 6302-2-1 &150 $\times$ 275 &  &  \text{\citet{2001M&PS...36..975H}}  & Compact CAI, lens shaped\\
Quinyambie 6076-5-1 & 50 $\times$ 120 & Type A & \citet{1996Sci...273..757R}  & Fragmented \\
\noalign{\smallskip}
\hline
\noalign{\smallskip}
Moo 1 & 350 $\times$ 150 & & \text{\citet{2018E&PSL.498..257E}} & Heavily altered, unknown precursor \\
Moo 2 & 240 $\times$ 260 & & \text{\citet{2018E&PSL.498..257E}} & Heavily altered, possibly type B precursor \\
HaH 1 & 130 $\times$ 90 & & \text{\citet{2018E&PSL.498..257E}} & Heavily altered, unknown precursor \\
HaH 2 & 190 $\times$ 160 & & \text{\citet{2018E&PSL.498..257E}} & Heavily altered, unknown precursor \\
Sah 041 & 300 $\times$ 140 & & \text{\citet{2018E&PSL.498..257E}} & Heavily altered, possibly type B precursor \\
Sah 196 & 100 $\times$ 80 & & \text{\citet{2018E&PSL.498..257E}} & Heavily altered, possibly type B precursor \\
\noalign{\smallskip}
\hline
\noalign{\smallskip}
NWA 5697-1 & 225 $\times$ 180 & Compact type A & This work & Melilite dominated \\ 
NWA 5697-2 & 50 $\times$ 40 & Fluffy type A & This work & Heavily altered, contains anorthite and fassaitic pyroxene \\
NWA 5697-3 & 30 $\times$ 25 & Pyroxene-hibonite & This work & Hibonite-pyroxene spherule with Ti-oxide\\
NWA 5697-4 & 30 $\times$ 25 &   & This work & Pyroxenitic fragment \\ 
\noalign{\smallskip}
\hline
\end{tabular}
\end{table*}

\begin{figure*}[h]
        \centering
        \includegraphics[width=0.38\textwidth]{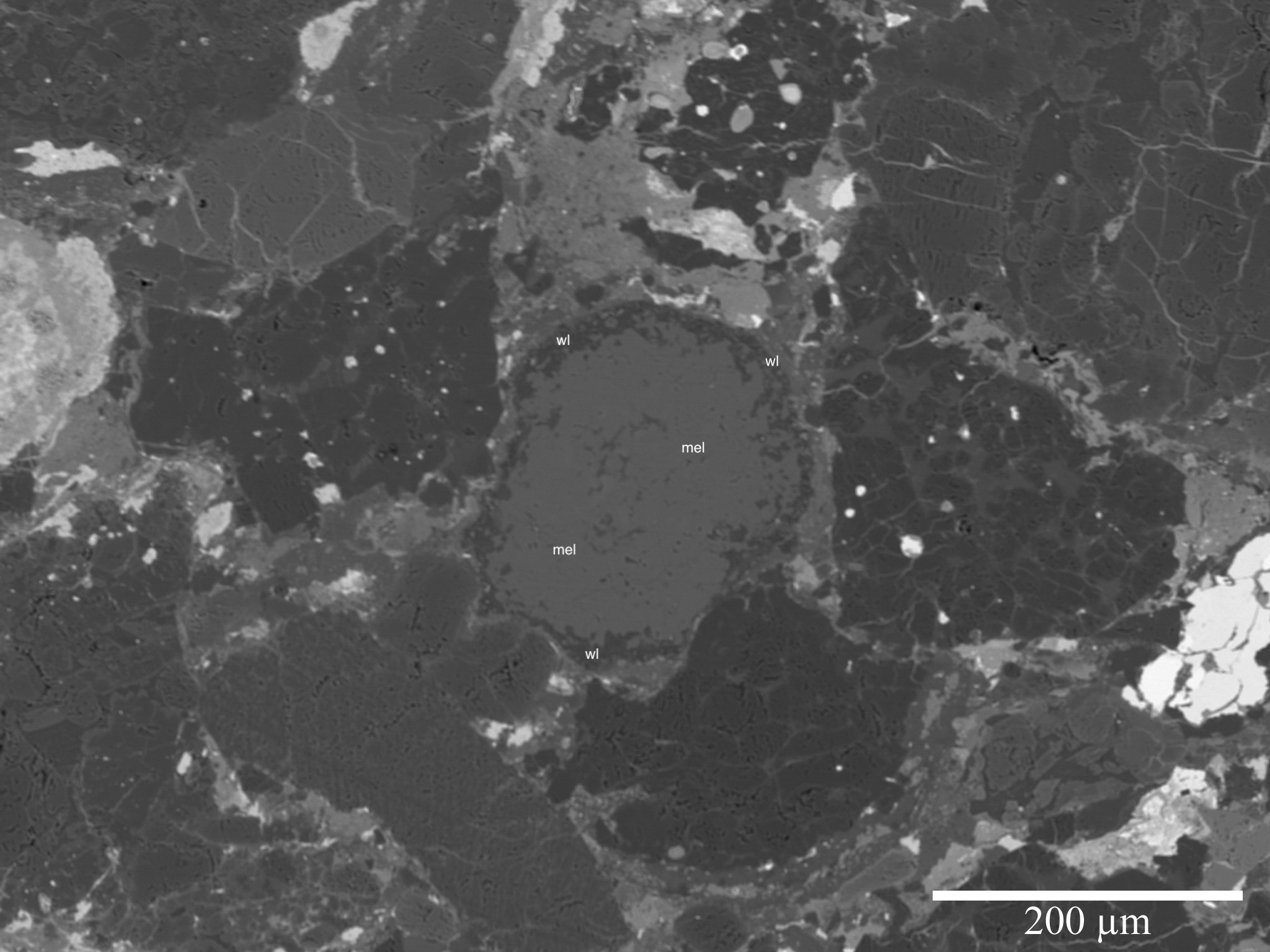}
        \includegraphics[width=0.38\textwidth]{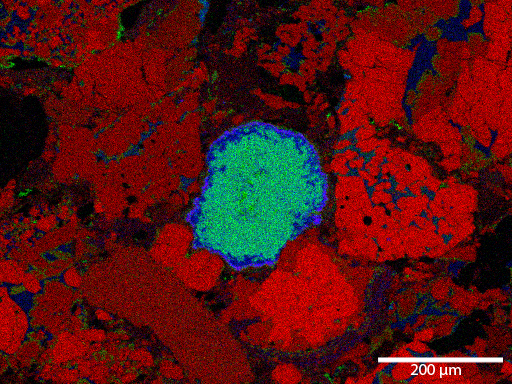}\\
        \includegraphics[width=0.38\textwidth]{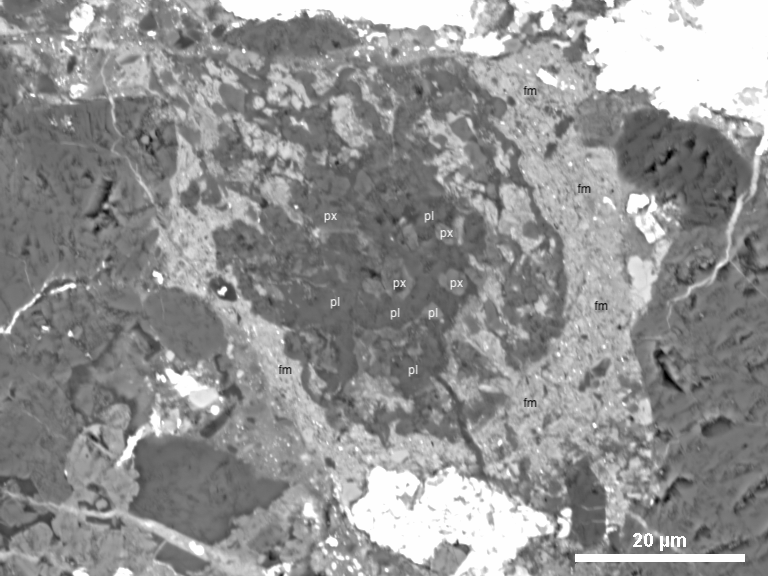}
        \includegraphics[width=0.38\textwidth]{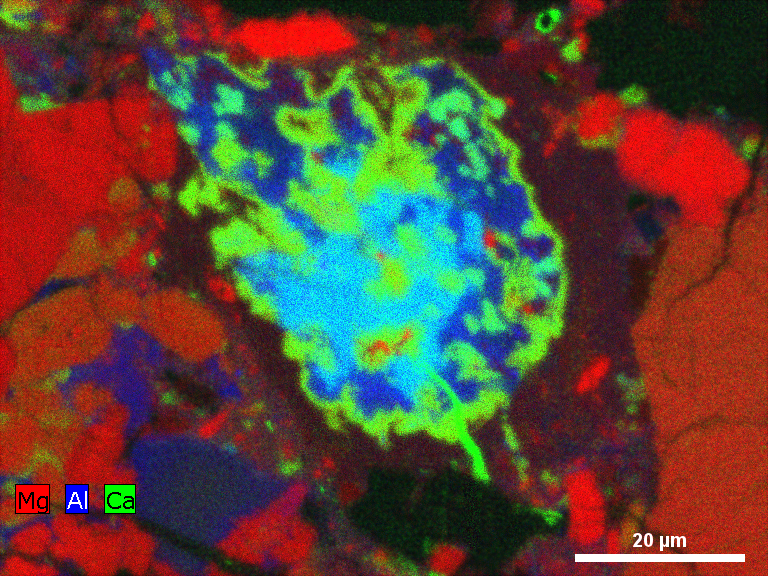}\\
        \includegraphics[width=0.38\textwidth]{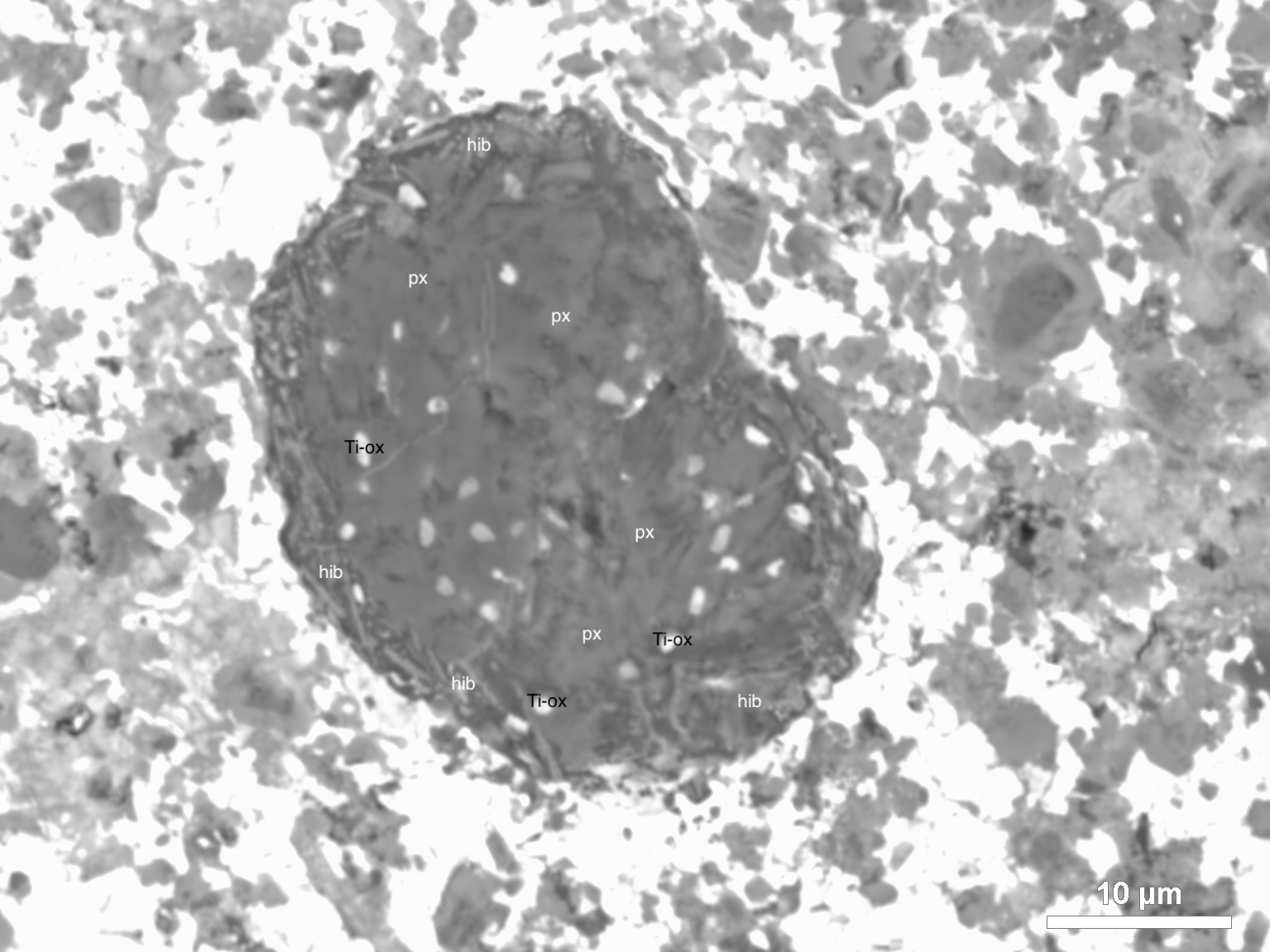}
        \includegraphics[width=0.38\textwidth]{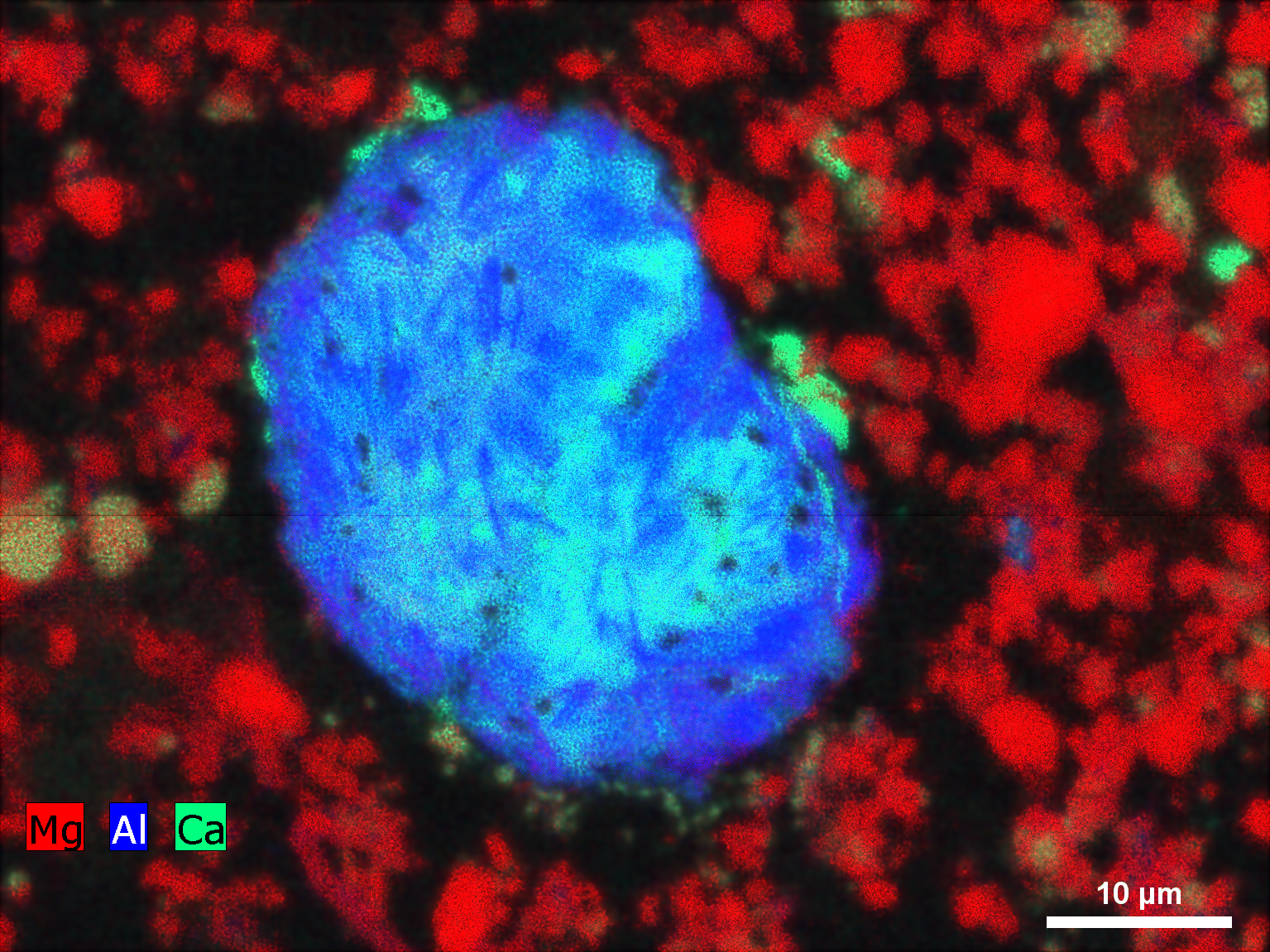}\\
        \includegraphics[width=0.38\textwidth]{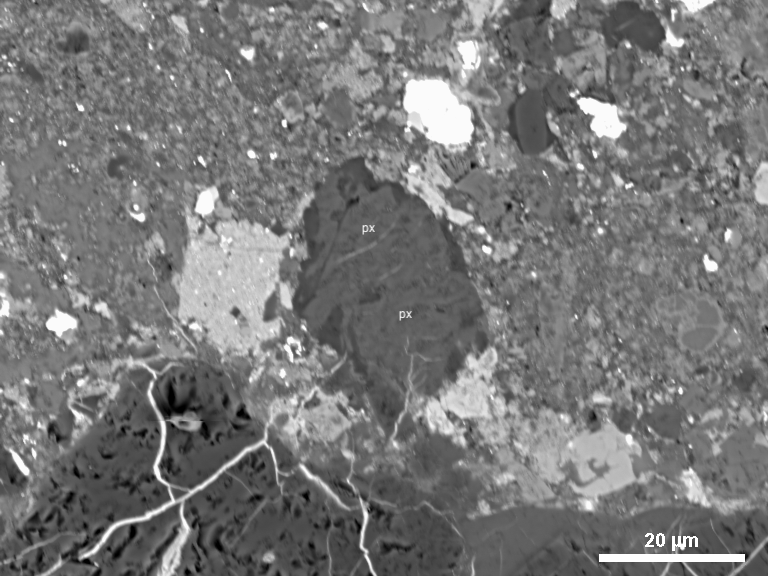}
        \includegraphics[width=0.38\textwidth]{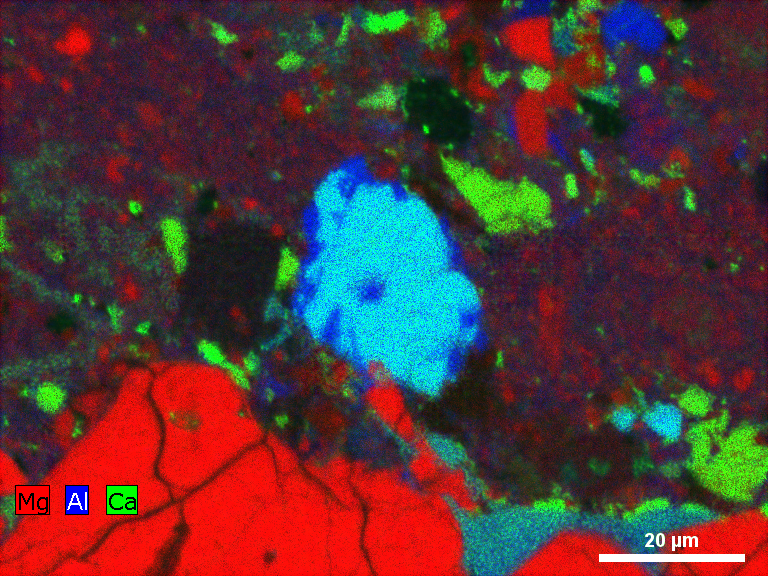}\\
        \caption{Backscatter electron images (left) and magnesium (red), calcium (green), and aluminum (blue) elemental maps (right) for
        NWA 5697-1 (top row), NWA 5697-2 (second row), NWA 5697-3 (third row), and NWA 5697-4 (bottom row). NWA 5697-1 is a
        Compact Type A composed primarily of Gehlenitic mellilite (mel) with a thin intact multi-mineralic Wark-Lowering rim (wl).
        NWA 5697-2 contains anorthitic plagioclase (pl) and fassaitic pyroxene (px), which seems to be embayed and/or replaced by
        fine-grained, more iron-rich material (fm).
	Given the convoluted texture, it could originally have been a fluffy type A. NWA 5697-3 consists primarily of hibonite (hib) and
	fassaitic pyroxene (px), with a significant component of Ti-oxide(s) of uncertain mineralogy.
	NWA 5697-4 seems to be composed primarily of fassaitic pyroxene (px).}
        \label{fig:scans}
\end{figure*}

Figure~\ref{fig:filter} illustrates the depletion factor $\zeta$ for five different $\alpha$-values. The triangles display the results of the simulations, to which we fitted a function of the form
\begin{equation}
        \zeta(a) = A\times \mathrm{erf}[B\times(a-C)]+D\, ,
        \label{eq:fit}
\end{equation}
where $\mathrm{erf}$ is the error function defined by
\begin{equation}
\mathrm{erf}(x) \equiv \frac{1}{\sqrt{\pi}}\int_{-x}^{x}e^{-b^2}db.
\end{equation}
The parameters $[A,B,C,D]$ in Equation (\ref{eq:fit}) are different for each level of viscosity and planet setup and given in Table~\ref{tab:fit}. For the large
majority of models, dust is efficiently filtered at large sizes, which implies that $\zeta$ vanishes for a large grain size, and therefore $D=-A$.

The depletion factor in Figure~\ref{fig:filter} shows  that with higher viscosity the filtering is shifted toward larger grain sizes, allowing more and larger dust grains to be transported to the inner solar system. The reason for this is twofold. First, in the case of a higher viscosity, the gap is refilled by gas faster, making it less pronounced. This benefits transport of the dust toward the inner system, since the pressure gradient at the outer gap edge is less steep and the gas surface density in the gap area is higher, mapping even larger dust grains to a small Stokes number according to Equation~(\ref{equ:Stokes}).
The second reason is that in our consideration the coefficient of dust diffusion $D_\mathrm{d}$ is set equal to the gas viscosity. Hence, by increasing the viscosity of gas we also increase the diffusivity of dust, also promoting dust transport to the depleted parts of the disk. In the extreme case of having no dust diffusion at all, the transport of dust is considerably less efficient, as has been studied in \citet{2018ApJ...854..153W}.

\subsubsection{Jupiter and Saturn acting in concert}
An interesting modification of the previous setup is to include Saturn alongside Jupiter.
Figure~\ref{fig:js-jsr} shows the resulting density profiles for the two different Jupiter--Saturn configurations in the fiducial disk model described in section~\ref{subsec:setups}. In the case of Jupiter and Saturn being at their current orbits (Setup II), the gaps of the two planets are separated and the filtration works analogous to a connection in series. The pressure gradient and related gap created by Saturn do not efficiently trap any of the regarded dust grains, but the more pronounced gap of Jupiter traps grains larger than $a\approx 1 \mathrm{cm}$ from the inner region. Grains with a size of $a\lessapprox 1 \mathrm{mm}$ are not as strongly affected by the filtration process, similar to Setup I.

In the case of Jupiter and Saturn entering a 3:2 MMR (Setup III), their gaps cannot be regarded separately anymore. The right panel of Figure~\ref{fig:js-jsr} shows that the two gaps overlap and that the presence of Saturn diminishes the overdensity that is created just outside of Jupiter's gap. An immediate consequence of this is that the pressure gradient at the outer edge of Jupiter's and Saturn's shared gap is less steep than in the case of an isolated Jupiter, making the filtration less efficient \rev{than in both previous setups} and therefore permitting larger dust grains to be transported to the inner system. This becomes even more apparent when comparing the left and the right panels of Figure~\ref{fig:js-filter} where the filtering functions $\zeta(a)$ of Setups II and III are displayed. The plots show that for any given $\alpha$-value the filter is shifted by about one order of magnitude to larger grain sizes in the scenario of Setup III. For the highest assumed viscosity $\alpha=10^{-2}$ no complete filtration was reached for the considered range of grain sizes. The reason for this is that the produced pressure profile is not able to efficiently prevent grains from diffusing through the gap owing to the large diffusion coefficient $D_\mathrm{d}$. The assumption of $D_\mathrm{d}= \nu$, however, could be an overestimate for large grains that reach $\St \gtrsim 1$ at the planetary gap. \citet{2007Icar..192..588Y} have argued that in those cases the diffusivity should be modified to $D_\mathrm{d}= \nu / (1 + \St^2)$, which would reduce the permeability of the gap for those grains.

\subsubsection{Model comparison \rev{and critical grain size}}
The grain size dependency for the transport of dust has been presented in the context of three different planetary configurations.
To describe the dependence of the filtration on the model parameters in a more quantitative manner and be able to compare
the setups with laboratory data, we introduce a critical grain size $a_\textrm{crit}$ above which the gap is filtering dust efficiently:
\begin{equation}\label{eq:acrit}
        \zeta(a_\mathrm{crit}) = 0.1\times \zeta_\mathrm{max}(a)
\end{equation}
In other words, we define the critical dust grain size as the size for which the transport becomes only 10\% as efficient as for the most coupled dust species.
We note that this convenient definition is purely by choice and the threshold value $0.1$ has no specific physical significance\rev{, but as can be seen in
figure \ref{fig:js-filter} the filtering is an exponentially steep function of size. Therefore, changing this definition by using, e.g.,~0.01 instead of 0.1 will increase
the critical size by less than a factor of two.}

Figure~\ref{fig:js-alpha} shows the dependence of the critical grain size on the viscosity of the gas. By fitting a power law to these values, we obtain a function for the critical grain size, in terms of $\alpha$, for the three different setups:
\begin{eqnarray}\label{equ:acrit1}
     \textrm{Setup\ I:} \qquad   a_\mathrm{crit}(\alpha) &=& 189\times\left(\frac{\alpha}{10^{-3}}\right)^{1.70}\,\mu\textrm{m}\\ [5pt]
\label{equ:acrit2}
\textrm{Setup\ II:} \qquad a_\mathrm{crit}(\alpha) &=& 740\times\left(\frac{\alpha}{10^{-3}}\right)^{1.00}\,\mu\textrm{m}\\ [5pt]
\label{equ:acrit3}
\textrm{Setup\ III:} \qquad a_\mathrm{crit}(\alpha) &=& 4086\times\left(\frac{\alpha}{10^{-3}}\right)^{0.93}\,\mu\textrm{m}
\end{eqnarray}
While \rev{there is only} one planet in Setup I (and therefore a completely autonomous gap) and a shared gap in Setup III, the nature of the gaps in Setup II depends on the level of disk viscosity. In the case of a high viscosity ($\alpha=10^{-2}$) the same result as for an isolated Jupiter is obtained, given that the potential of Saturn is perturbing the disk only very locally. For low viscosities, however, the filtration is less efficient than in Setup I and the $\alpha$-dependence in Figure~\ref{fig:js-alpha} resembles more the slope of Setup III, where the planets share a gap.  \citet{2016PASJ...68...43K} investigated the dependency of the width of a planetary gap, $\Delta_\mathrm{gap}$, on the planet's mass and location and the disk parameters. They found from numerical studies that
\begin{equation}\label{eq:width}
        \Delta_\mathrm{gap} = 0.41  \left(\frac{M_\mathrm{P}}{M_\ast}\right)^{1/2}h^{-3/4}\alpha^{-1/4} r_\mathrm{P}\,.
\end{equation}
This shows that by reducing the viscosity the gap structure grows not only in depth but also in width. In case of a high viscosity the gaps of Jupiter and Saturn are rather narrow and do not influence each other. The filtration imposed by Jupiter is stronger than Saturn's, and therefore the same value of $\zeta$ as for the isolated Jupiter is found. For low viscosities, the gaps grow in width, start to overlap, and consequently the transport toward the inner system is enhanced compared to the stand-alone Jupiter scenario. The critical value obtained from Equation~(\ref{eq:width}) is $\alpha \approx 5\times 10^{-4}$, but as Figure~\ref{fig:js-alpha} shows, the transition between the two regimes is not abrupt but continuous.

\rev{The critical grain size is useful because it compares directly to the upper size limit of CAIs found in chondrites that accreted in the inner solar system.
The model dependence of the critical size allows us therefore to constrain the physical conditions in the accretion disk and the planetary configuration in the
vicinity of Jupiter.}

\subsection{Refractory inclusions in ordinary chondrites}
The result of the systematic scan of the slabs from the NWA 5697 is reported in table \ref{tab:caifrags}
together with refractory inclusions in OCs known from literature
\citep{1984GeCoA..48..693B,2001M&PS...36..975H,1996Sci...273..757R,2016LPI....47.1989R,2018E&PSL.498..257E}.
\rev{An additional collection of 24 CAIs in OCs was reported by \citet{lin2006petrographic}. We chose not to include this dataset owing to
the heavily altered condition of the inclusions and the resulting risk of including false positives in the form of, e.g., Al-rich chondrules
in the combined dataset. After consideration and follow-up measurements of more than 100 candidates,} we have identified
one large CAI and three additional small CAIs.
BSE images and magnesium--calcium--aluminum maps of the four identified CAIs are shown in figure \ref{fig:scans}.

With the $\mu$-XRF scanner we surveyed 53 cm$^2$ in three polished slabs from the OC
NWA 5697 and found one large CAI
with an area of 0.0004 cm$^2$ corresponding to a fractional areal density of CAI material in OC slabs of
$\delta_\textrm{CAI, in} = 7.5 \times 10^{-6}$, in accordance with
other estimates of an areal density of less than 10$^{-4}$ \citep{2005ASPC..341...15S}. A similar estimate can be made based on the 4.1 cm$^2$
surveyed with the SEM. Here the ratio becomes $8.5  \times 10^{-6}$, which is in accordance
with the large-scale scan. 
Given that only four CAIs are identified, it is not possible to quantify a size distribution, but the sizes and relative abundance are
consistent with what is found in CV / CK chondrites, if CAIs larger than $\sim$200 $\mu$m are disregarded \rev{in the CV / CK chondrite distribution}
\citep[see, e.g., ][]{2008M&PS...43.1879H}.

The largest of the CAIs, NWA 5697-1, is a compact type A, which is already known from other OCs (see table \ref{tab:caifrags}).
While it is possible that the
unaltered precursor of CAI NWA 5697-2 was in fact a Type B or C igneous CAI, the first such CAI recognized in OCs,
the convoluted texture rather resembles fine-grained CAI, which is our preferred interpretation. The pyroxene-hibonite spherule NWA 5697-3
is, according to the compilation of \citet{2014mcp..book...65S}, the first CAI of its kind recognized in OCs, and as such it extends
the repertoire of these meteorites. As such, the main \rev{type in the carbonaceous chondrite CAI inventory not found yet in OCs} is the type B CAI that
typifies CV chondrites and coincidentally the many isotopic studies that rely on their typical large sizes and well-defined mineralogy.
Whether this absence is a result of the suppression of large CAIs in OCs or the current small number statistics is uncertain and a question
that has to be resolved by a larger-scale follow-up study.

\rev{After the submission of this study, \citet{2018E&PSL.498..257E} published a new compilation of six CAIs in OCs with a geometrically averaged linear
size ranging from 90 $\mu$m to 250 $\mu$ sampled from three different OCs (HaH 335, Moorabie, Sahara 98175) supporting the maximum CAI size
found in NWA 5697.}

\section{Discussion} \label{sec:discussion}
In this section we link the results that we found from numerical models to the obtained cosmochemical evidence. We also outline possible shortcomings and how they may affect the obtained results.

\subsection{Constraining models with cosmochemical data}
\begin{figure}
        \centering
        \includegraphics[width=.44\textwidth]{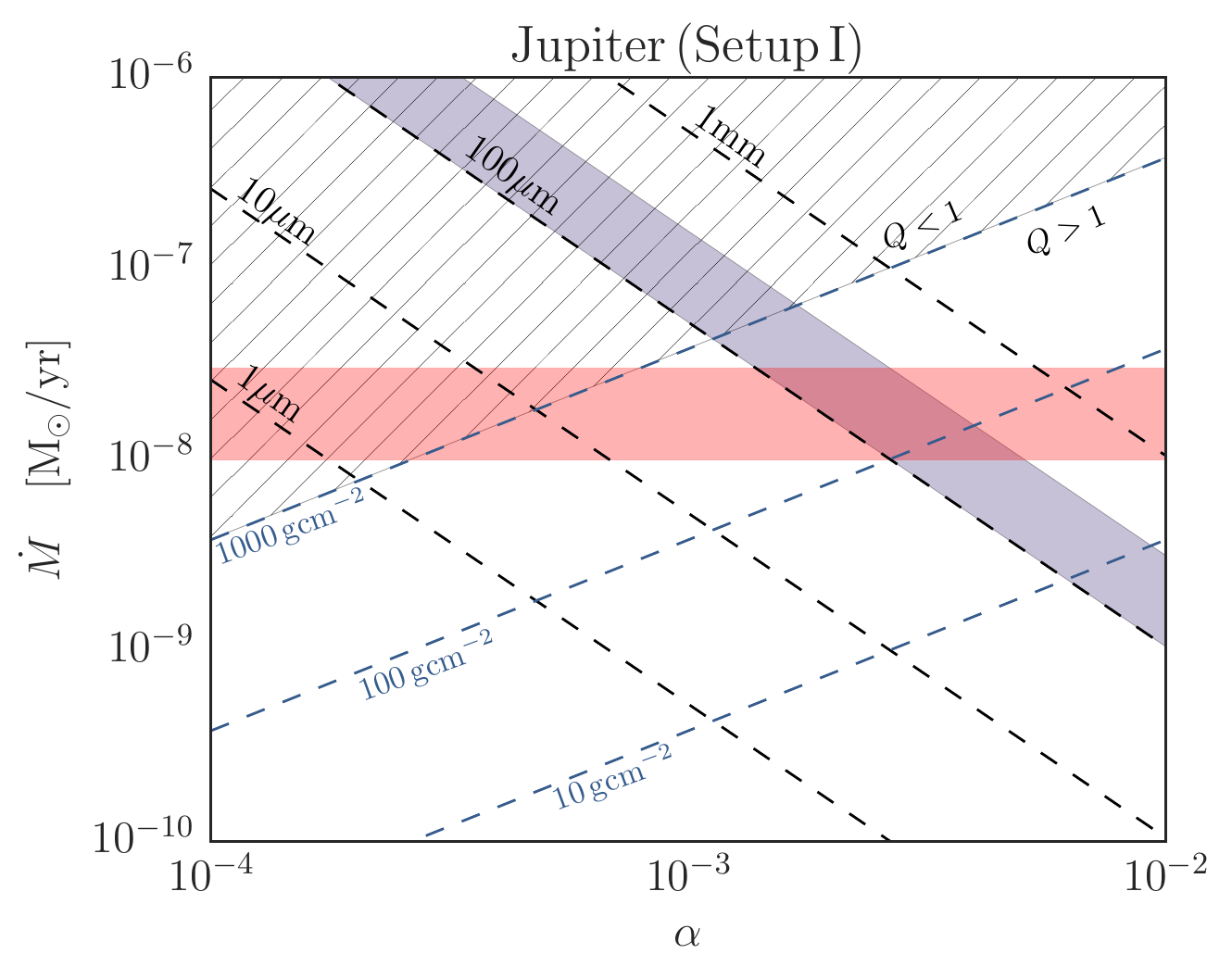}\\[-12pt]
        \includegraphics[width=.44\textwidth]{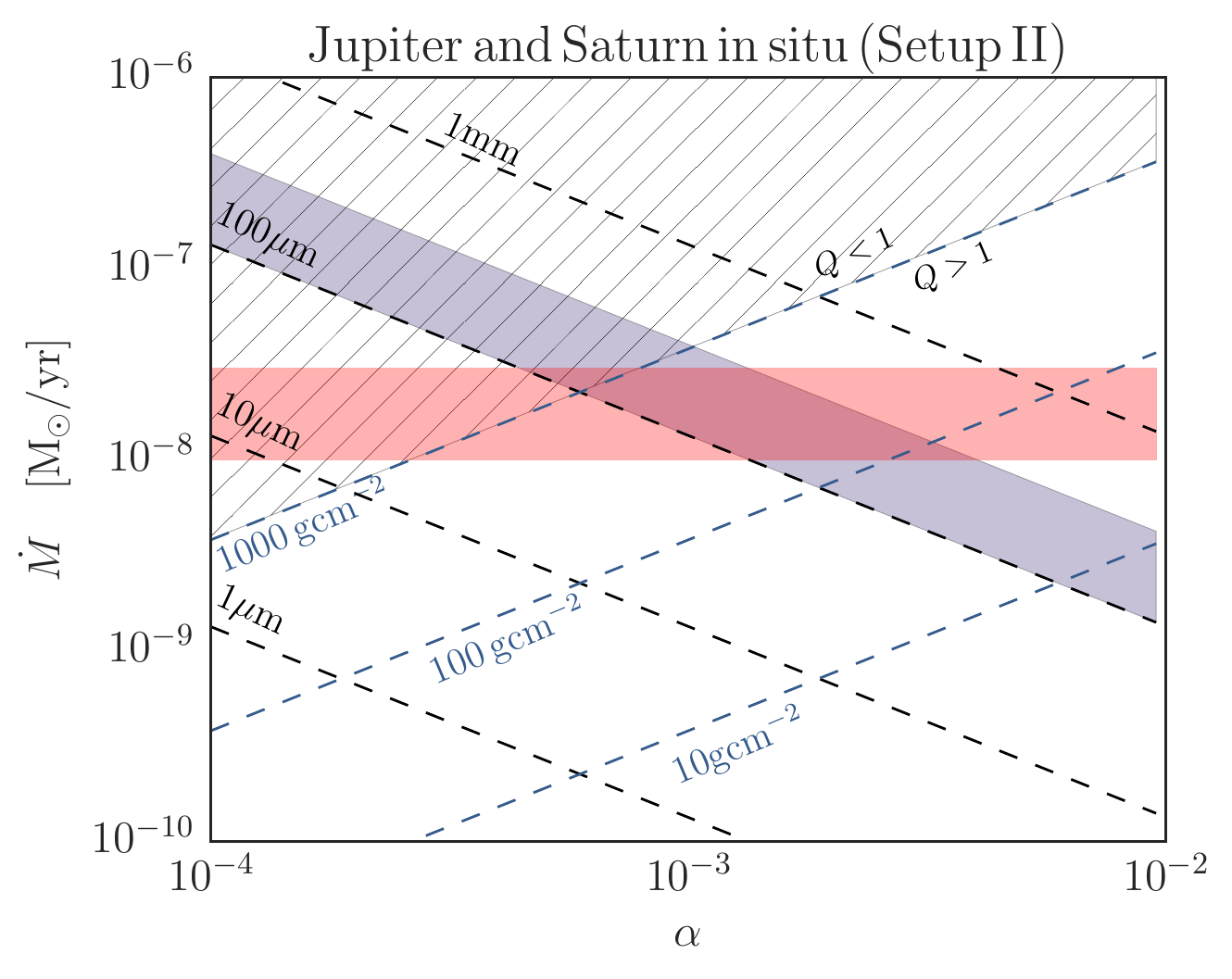}\\[-12pt]
        \includegraphics[width=.44\textwidth]{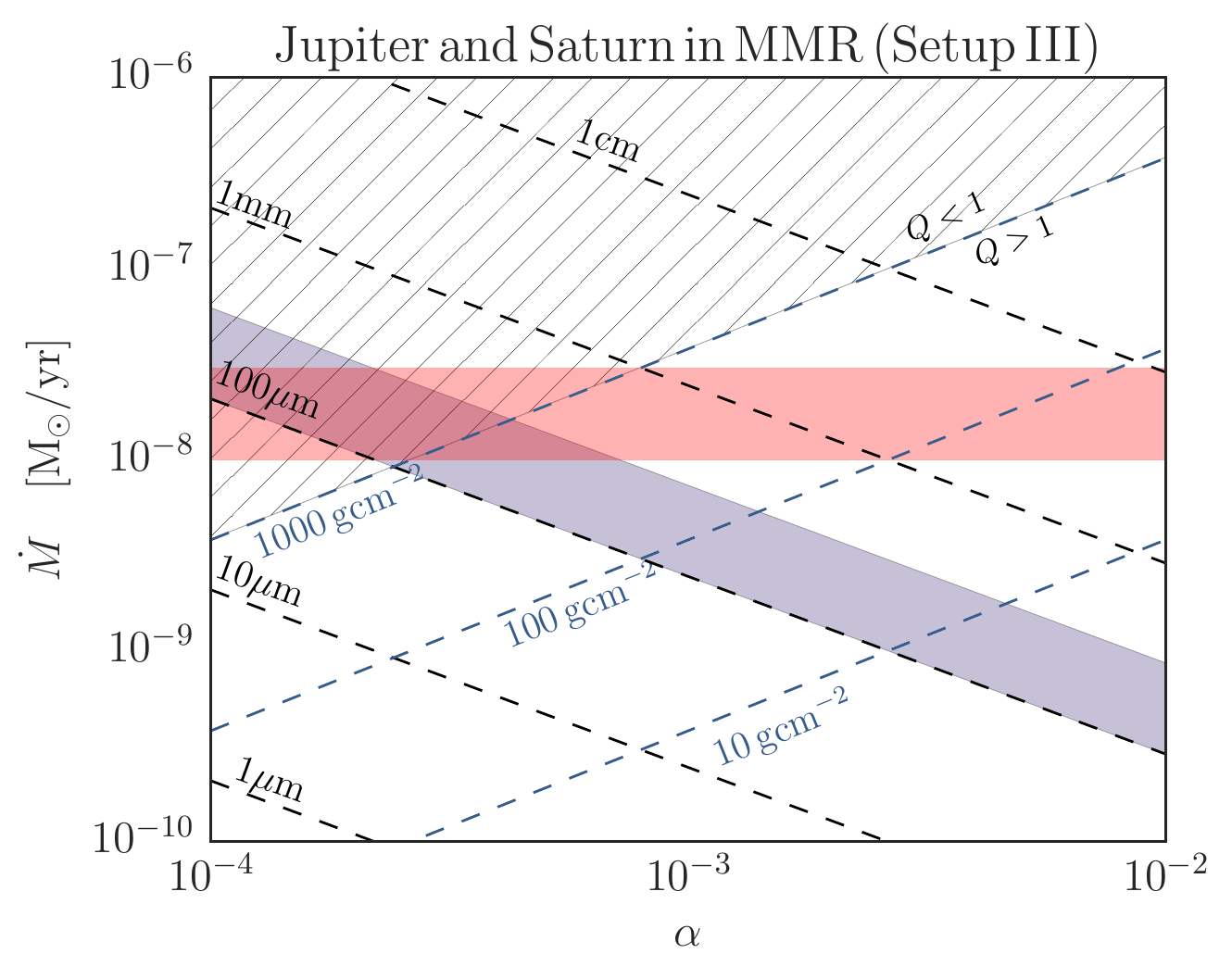}
        \caption{Critical grain size as a function of model parameters. Downward dashed black lines show the critical grain sizes obtained from the numerical models for the case of Jupiter (Setup I: top panel), Jupiter and Saturn in situ (Setup II: middle panel), and Jupiter and Saturn in MMR (Setup III: bottom panel).
        Light-blue shaded areas highlight the critical size estimated from the laboratory measurements, and diagonally upward dashed light-blue lines indicate the unperturbed surface densities at the location of Jupiter. Light-red shaded areas show the assumed maximum stellar accretion rate during the epoch of free-streaming CAIs outside Jupiter.  }
        \label{fig:colorfilter}
\end{figure}

Equations~(\ref{equ:acrit1}) -- (\ref{equ:acrit3}) parameterize the critical sizes above which the grains are blocked efficiently for entering the
inner solar system for the three different planet configurations. They are based on simulations that all use a stellar accretion rate of
$\dot{M} = 10^{-7}\,\mathrm{M}_\odot\,\mathrm{yr}^{-1}$. In the modeling framework
viscosity and the stellar accretion rate are kept as input parameters. Together they
determine the absolute value of the gas surface density. Assuming a different stellar accretion rate as an input parameter therefore modifies the
the gas surface density, which changes the coupling between dust and gas characterized by the Stokes number
$\St \propto a\, \Sigma_\mathrm{g}^{-1}$ (see Equation~(\ref{equ:Stokes})). To keep the Stokes number and therefore the numerical results
invariant, a change in $\Sigma_\mathrm{g}$ results in a corresponding change in the grain size $a$. Using the linear
$\Sigma_\mathrm{g}$ --  $\dot{M}$ relation, we can express the change in critical grain size as a function of the assumed stellar accretion
rate and by that extend the parameter space of $a_\mathrm{crit}(\alpha)$ in Equations~(\ref{equ:acrit1}) - (\ref{equ:acrit3}) as
\begin{equation}
        a_\textrm{crit}(\alpha,\dot{M}) = \left( \frac{\dot{M}}{10^{-7}\,\mathrm{M}_\odot\,\mathrm{yr}^{-1}}\right)\,a_\textrm{crit}(\alpha)\,.
        \label{equ:amdot}
\end{equation}
This scale-free critical dust grain size is illustrated by black dashed lines in the three panels of Figure~\ref{fig:colorfilter}.

\subsubsection{Critical grain size and assumed stellar accretion rate}
In Table~\ref{tab:caifrags}, we presented cosmochemical measurements of fragments of CAIs. These data and the assumption
that CAIs followed the inflow of material in the envelope located outside of Jupiter's orbit, at a time when Jupiter was already present,
suggest that the critical grain size is $a_\mathrm{crit} = 100\textrm{ -- }300 \,\mathrm{\mu m}$.

Thirteen individual chondrules in OC NWA 5697 have been dated using U--Pb absolute dating and record a time span covering from 0 to 4 Myr
after the formation of the solar system \citep{2017SciA....3E0407B}. Either the core of the parent chondrite body was built
up fast, and then a thin surface layer of material was added over 4 Myr, or chondrules were efficiently stored in the inner solar system for
extended time periods and the parent body was only formed after 4 Myr. In either case, the lack of refractory inclusions, compared to the
prevalence in CV and CK chondrites, indicates that Jupiter and/or Saturn acted as an effective barrier over a long period of time.

\rev{As discussed below in section \ref{sec:dustdistribution}, if larger CAIs present in the CV / CK accretion region were allowed to enter the accretion
region of NWA 5697 not all would be lost through reprocessing of material, and we would have readily identified them in our lower-resolution, larger-area
scan. The absence of large CAIs therefore has to be the result of active filtering and not reprocessing. The exact amount of reprocessing
could shift the critical size, but the function is steeply declining (see Fig.~\ref{fig:filter}) and not too sensitive to the exact value of the critical size.
While a larger statistical sample is desirable, the absence of large CAIs makes $a_\mathrm{crit} = 100\textrm{ -- }300 \,\mathrm{\mu m}$ a
reasonable estimate, given the sampling uncertainties discussed below.}

Evidence from observations and models of star forming regions suggests that typical stellar accretion rates during this time span
could be anywhere from 10$^{-5}$ to 10$^{-9}\,\mathrm{M}_\odot \mathrm{yr}^{-1}$, with the accretion being toward lower values at later times.
If the core of proto-Jupiter was in place at the end of the Class I phase, where the last of the envelope, together with the CAIs, is being accreted,
this range of accretion rates is presumably lower and would be in the 10$^{-8}\,\mathrm{M}_\odot \mathrm{yr}^{-1}$ range.

\subsubsection{Exploring different planetary configurations}
The purpose of Figure~\ref{fig:colorfilter} is to directly compare the obtained results from different numerical models to the value of
$a_\mathrm{crit}$ suggested by the cosmochemical measurements. 
The plots show the critical grain size and the initial gas surface density at Jupiter's location, $\Sigma_0$, for combinations of
$\alpha$ and $\dot{M}$ according to Equation~(\ref{equ:amdot}). 
The blue shaded area incorporates the estimate of $a_\mathrm{crit}$ for CAIs of 100--300$\mu$m,
while the red shaded area indicates a range for the maximal accretion rate, $\dot{M} = 10^{-8}\textrm{ -- }3\times10^{-8}$, during the
epoch where CAIs were accumulating outside Jupiter in the protosolar accretion disk.
In our numerical models the gas disk is not self-gravitating, and we therefore implicitly assume the disk to be gravitationally stable for the
entire computational domain. The hatched part of the  parameter space is where the Toomre parameter,
$Q \equiv c_\mathrm{s}\Omega_\mathrm{K}/(\pi G \Sigma)$, becomes smaller than unity -- more specifically, where $Q\!<\!1$ at
some point inside 16 au, which is the upper boundary of the numerical disk model.

We now examine the implications for different planetary configurations in the context of the results presented in Figure~\ref{fig:colorfilter}.
The top panel shows the case of Jupiter acting as the only perturber in the disk (Setup I). The measurements of the critical grain
size and the assumption of the stellar accretion rate mentioned above limit the viscosity to $2\times10^{-3}\lesssim\alpha\lesssim 5\times 10^{-3}$
and suggest an initial gas surface density at Jupiter's location of $50\textrm{ -- }600\,\mathrm{g}\,\mathrm{cm}^{-2}$. This is consistent with
upper limits for $\alpha$ found in accreting protoplanetary disks \citep[e.g.][]{2017ApJ...843..150F,2018ApJ...856..117F} and also with the
gas surface density level proposed by the model of the Minimum Mass Solar Nebula \citep[MMSN; ][]{1981PThPS..70...35H}.
The assumption that CAIs only return to the inner solar system at a later time when the stellar accretion rate has already dropped
to lower values demands an even higher level of viscosity.

Adding Saturn to the setup at the location where it is today (Setup II; middle panel of Figure~\ref{fig:colorfilter}) leads to similar results to those in Setup I.
The proposed value  of $\alpha$ decreases and centers on $6\times10^{-4}\lesssim\alpha\lesssim 4\times 10^{-3}$ and this planetary configuration allows
$\Sigma_0 = 70\textrm{--}1000 \,\mathrm{g}\,\mathrm{cm}^{-2}$.
With Jupiter and Saturn in a 3:2 MMR (Setup III; bottom panel of Figure~\ref{fig:colorfilter}), a critical grain size of
100--300 $\mu$m can only be accommodated by smaller stellar accretion rates, $\dot{M}$, than indicated by the red shaded area.
In the classical picture, where the stellar accretion rate is expected to decrease with time, this would suggest that the filtering took
place at a later stage, where $\dot{M}$ had already reached smaller values such as $\dot{M}\simeq 10^{-9}\, \mathrm{M}_\odot \mathrm{yr}^{-1}$.
This is only possible if CAIs returned to the inner solar system at very late times, which is inconsistent with the protostellar envelope
almost having disappeared at the beginning of the T Tauri stage of $\approx$1 Myr. It is also inconsistent with any prolonged episodic accretion
events occurring at later times. Such events temporarily increase
the accretion rate and viscosity, which would allow outer solar system material with larger grain sizes to pass through the gap.
Therefore, this scenario would require a fine-tuned evolution of the young solar system with a large and low-density envelope where most
of the CAI material was dispersed to the outer edges, as well as a
prolonged period of low levels of accretion that proceeded in a smooth manner, suggesting a quiet low-density birth environment in a dispersed star
forming environment comparable to, e.g., present-day Taurus. This is at odds with the high levels of short-lived radio nuclides, such as
$^{26}$Al inferred to be present at the birth of the solar system, which requires a larger and
denser star forming environment \citep{2016ApJ...826...22K}.

\subsection{Overall distribution of dust} \label{sec:dustdistribution}
  
The critical size of CAIs in OCs gives clues to the overall dynamical properties in the protosolar disk, but in addition, the
areal density of CAIs inside and outside of Jupiter can inform us about the relative dust masses in the two regions, if we make a number of
reasonable assumptions. The fractional areal density of CAIs in CK and CV chondrites is $\delta_\textrm{CAI, out} = 4\%\textrm{ -- }10\%$
\citep{2008M&PS...43.1879H}.
The large CAIs that we find in OCs have to be just at the turnover in the depletion parameter, because otherwise we would find larger CAIs,
and on the other hand, if they were larger than the turnover, we would not find any, given that the transition is relatively sharp. Therefore, the depletion
parameter for the large CAIs has to be $\zeta_\textrm{CAI} = 0.01\textrm{ -- }0.001$. Another factor that affects the estimate is reprocessing. There is
evidence that reprocessing was widespread in the inner solar system from the abundance of aluminum-rich chondrules that to some extent reflect
remolten CAI material \citep{krot+2010}, the high
occurrence of chondrule fragments \citep{2005ApJ...623..571S}, and that younger chondrules have evolved initial Pb isotope compositions
\citep{2017SciA....3E0407B}. \rev{In addition, parent body impact heating preferentially heats smaller grains and the matrix reaching localized peak
temperatures in the fine-grained material of more than 2000 K for realistic impact velocities \citep{2014NatCo...5E5451B}}. This could lead to additional in situ reprocessing of small
CAI material on the parent body. It is difficult to estimate the fraction of large CAIs that survive intact, $f_\textrm{CAI}$, before incorporation in a parent body.
This cannot simply be estimated based
on the fraction of old and young chondrules in, e.g., NWA 5967, since the relative production rate or amount of raw material available is unconstrained,
as is the point in time when CAIs reached the region close to Jupiter. \rev{An estimate can be obtained by comparing the fraction of bulk CAI-like  
material in OCs ($< 2\times10^{-3}$)  to the fractional areal density of intact CAIs
($\delta_\textrm{CAI, in} = 10^{-5}$), suggesting that the survival rate is $f_\textrm{CAI} \approx 0.01$. We computed the upper limit for the pollution by CAIs
to the bulk mass of OCs by calculating the maximum (2$\sigma$) deviation of $\epsilon^{50}$Ti in the OC Saint-Severin from the
$\epsilon^{50}$Ti to the $^{54}$Cr--$^{50}$Ti mixing line reported in \citep{2009Sci...324..374T}.}
The relative masses of the dust reservoirs
inside and outside of Jupiter, $M_\textrm{in}$ and $M_\textrm{out}$, can now be calculated as
\begin{equation}
M_\textrm{in}\,\delta_\textrm{CAI, in} = M_\textrm{out}\,\zeta_\textrm{CAI}\,\delta_\textrm{CAI, out}\,f_\textrm{CAI}\,,
\end{equation}
which can be rewritten with probable values as
\begin{equation}
\frac{M_\textrm{in}}{M_\textrm{out}} = 0.5 \left(\frac{\zeta_\textrm{CAI}}{0.01}\right) \left(\frac{\delta_\textrm{CAI, out}}{0.05}\right)
    \left(\frac{f_\textrm{CAI}}{0.01}\right) \left(\frac{\delta_\textrm{CAI, in}}{10^{-5}}\right)^{-1}\,.
\end{equation}
Notice that this equation is only applicable in a steady-state flow where sinks (planetesimals and accretion flow) and sources (accretion from outside)
are more or less balanced, \rev{and therefore the result is only an order-of-magnitude estimate of the dust reservoirs}. That there are more or less equal
amounts of dust just outside and inside of Jupiter indicates a relatively shallow surface density profile as used in the paper, inside the considerable
uncertainties in the estimate. It supports the overall theoretical framework of large-scale mass transport outward at early times and the filtration
of large grains in planetary gaps at late times.

\subsection{Limitations}
In the numerical model dust is treated as a fluid, and we simulated several dust species over many thousands of orbits. Still, not all relevant physical
effects are included, and there is also uncertainty about the detailed evolution history of the young solar system that is not englobed by the models.
The cosmochemistry data come with natural limitations in terms of sampling bias and statistical weight.
In the following, the most important shortcomings are shortly outlined and discussed.

\paragraph{Planetary growth and migration},---\rev{In the presented numerical studies we fixed the masses and orbital radii of Jupiter and Saturn in time and are therefore not accounting for planetary growth or considering planet migration scenarios. The qualitative effect of a growing planet can be seen in \citet{2018ApJ...854..153W}, where the permeability of the gap was studied for different planet masses, showing that the filtration becomes more efficient as the mass of the planet increases. Conversely, the filtration would have been less pronounced at times when Jupiter (and Saturn) were not yet fully developed. The treatment in this work implicitly assumes that both Jupiter and Saturn have already grown to their final masses at the moment when the filtration of CAIs becomes relevant. This is motivated by expected rapid time-scales of runaway gas accretion \citep[e.g.][]{1996Icar..124...62P,2010MNRAS.405.1227M} that suggest that the time span between a planetary core affecting the most susceptible dust particles and this core's growth to a gas giant planet is relatively short.
Planet migration, i.e., planets changing their orbital radius, is an important mechanism that might change the predicted filtration mechanism, but it is outside the scope of this work. Both \citet{2014ApJ...792L..10D} and \citet{2015A&A...574A..52D} investigated the migration rate of a massive planet in a viscous accretion disk, taking into account the substantial gas flow through the gap created by the planet. They found inward-directed migration for a single planet that is not locked to the viscous accretion speed of the gas. \citet{2015A&A...574A..52D} showed that a migrating planet deforms the gap structure in a way that the gas density profile at the outer gap edge becomes less steep compared to fixed orbit simulations, which would lead to a higher permeability for dust grains and shift the critical grain size of filtration to larger values. In the case of Setup III, where Jupiter and Saturn enter a 3:2 MMR and share a wide gap in the gas structure, the migration behavior is more complex. \citet{2001MNRAS.320L..55M} showed that in such a scenario the two planets can migrate outward \citet{2012ApJ...757...50D} pointed out that the rate and direction of this migration strongly depend on the assumed disk structure. The effect of an outward migration of Jupiter and Saturn locked in a 3:2 orbital resonance has been studied only recently by \citet{2019AJ....157...45M}. They find that the outward migration is promoting the transport of dust from the outer reservoir to the inner system, making the filtration less efficient. In summary, it is not yet well explored, but the literature indicates that dust filtering for migrating planets could be less efficient.}

\paragraph{Dust feedback and dust evolution},---As mentioned in section~\ref{subsec:gas} we do not consider the friction force feedback that is exerted by the dust onto the gas in our models. Treating dust as test fluids makes the results independent of the assumed grain density and size distribution, while it does not strictly conserve the total momentum. The approach is valid as long as the local gas density is considerably higher than the dust density. In pressure bumps, where dust can be trapped, it is not always the case. However, as discussed in \citet{2018ApJ...854..153W}, while the location of the outer dust density peak is at a slightly larger distance to the planet's orbit, the filtering of the dust at the gap remains unchanged when feedback is considered.\\ 
Further, dust grains can grow by coagulation and condensation or fragment and evaporate, replenishing a population of smaller grains, but
the effects of dust evolution are not included in our study. This is not a limitation for the comparison of the numerical models to the laboratory measurements, given that CAIs are
unprocessed and have not interacted with other grains; however, it demands caution when comparing the CAI size distribution to the size distribution of solids in general. Notice that in OCs there is ample evidence of reprocessing, and,
e.g., aluminum-rich chondrules could have been created by the reprocessing of refractory material, as discussed above.

\paragraph{Viscous disk evolution},---The $\alpha$-model has to be regarded as a gross simplification, and it moreover relies on a readily available,
robust source of turbulence within the disk. \citet{2014prpl.conf..411T} provide an account of potential alternative transport mechanisms.
More recently, \citet{2018arXiv180404265K} have explored the scattering of X-rays into gaps opened by Saturn- and Jupiter-mass planets and
arrive at the conclusion that the magnetorotational turbulence is an unlikely scenario given the resulting ionization state of the gap.
More probably, the Hall-shear instability \citep[HSI; ][]{Kunz2013a} may provide the means of accreting the gas in a laminar flow caused by
large-scale magnetic torques.
It is important to stress that transport of angular momentum by HSI or magnetocentrifugal winds may invalidate the direct connection
between $\alpha$, $\dot M$, and $\Sigma$ that we exploit in Equation (\ref{equ:gasdens}). If additional sources for transport of angular
momentum exist, it becomes an upper limit $\alpha < \dot M \Omega_\textrm{K} / (3\pi c_\textrm{s}^2 \Sigma_\textrm{g})$.
In magnetically decoupled regions, purely hydrodynamic instabilities
\citep[such as the vertical shear instability, VSI, ][]{2013MNRAS.435.2610N,2016A&A...586A..33U} may arise, but the redistribution
of embedded solids \citep{2016A&A...594A..57S}, may deviate from a purely diffusive process. Nevertheless, \citet{2018A&A...616A.116P} have
recently studied the evolution of particles in the vicinity of a planetary gap and have found that the behavior of particles is very similar for the case of a VSI turbulent disk and in laminar models with stochastic kicks.

\paragraph{Models of the solar nebula},---The slopes of the surface density and temperature are set up to produce a steady-state disk in the unperturbed limit with a radial dependence
of the surface density of $\Sigma_\mathrm{g} \propto r^{-0.5}$. In some solar system models it is argued  that this slope is much steeper.
One of the most prominent models is the MMSN, for which distributing the mass of the planets over the planetary disk leads to a slope of
$\Sigma_\mathrm{g}\propto r^{-1.5}$ \citep{MMSNWeidenschilling,1981PThPS..70...35H} and in a further study even
$\Sigma_\mathrm{g}\propto r^{-2.168}$ \citep{2007ApJ...671..878D}. Apart from changing the time-scales of dust grain drift and the viscous
disk evolution, this would be particularly relevant in models that consider Saturn alongside Jupiter, given that it produces a higher density
contrast between the location of Jupiter and Saturn. Consequently, there would be less relative gas density at Saturn's location than in our
model -- making the filtration by Saturn more efficient.\\
Nevertheless, we assume a rather flat pressure profile to obtain a steady-state solution in the unperturbed case, which is important to
implement adequate inflow and outflow values in our numerical studies. Additionally, some observations of gas tracers in protoplanetary
disks suggest that the slopes derived from the MMSN models are exaggerated \citep[e.g.][]{Williams2016}. 

\paragraph{Evolution of the protosolar disk, and the mass accretion of the gas giants},---The mass of the protosolar disk and the mass of the planets evolve over time. The change in mass affects the typical size of the grains
being filtered. Numerical models \citep{2014ApJ...797...32P,2016A&A...587A..59F,Kuffmeier+2017} and
observations \citep{2009ApJS..181..321E,2016A&A...594A..85M,2017A&A...599A.113M} of
class I protostars and pre-main-sequence stars show a wide variety in disk masses and accretion rates, but with an overall
decrease in both over time. As discussed above, we have tried to be conservative about the maximum size of the accretion
rate, when estimating the surface density at Jupiter's orbital location. On top of the secular trends, the existence of episodic accretion events around
other stars is now well established both observationally and in models 
\citep{1985ApJ...299..462H,2009ApJ...702L..27B,2012ApJ...756..118B,2014ApJ...797...32P,2017A&A...602A.120F,2017arXiv171202457J,2018MNRAS.474.1176J,2018MNRAS.475.2642K},
and it is reasonable to expect the young Sun to have had similar outbursts. These episodic accretion events simultaneously increase the stellar accretion rate
and the average temperature, due to the sharp increase in the luminosity, thereby increasing the viscosity in the disk.
The net effect will be a temporary decrease of the filtration efficiency and the allowance of larger dust grains to pass through the gap. It is an open problem,
though, whether an event lasting a decade or a century will be enough to have a significant effect on the total amount of CAIs,
or how many orbital times are necessary to effectively change the transport properties of the dust grains.
The evolution in the planetary masses will also affect the filtering of dust grains, as can be appreciated by comparing the different filtering efficiencies
of Jupiter and Saturn (e.g., Figure \ref{fig:js-alpha}). If CAIs had arrived close to Jupiter before it had built up a large fraction of its current mass,
CAIs larger than 100 $\mu$m should have had no problem in passing through the gap. The absence of such objects in OCs strongly suggests
that Jupiter formed early and / or CAIs arrived late. Notice, though, that because CAIs had to accrete as part of the envelope, the arrival of CAIs at
Jupiter cannot have happened much later than 1 Myr after the formation of the solar system.

\paragraph{Cosmochemical variance in OCs},---In our study we have made an unbiased serendipitous high-resolution SEM scan in four small areas of polished sections extracted from the OC
        NWA 5697 covering in total an area of 4.1 cm$^2$. This has allowed us to identify CAIs in the 10 -- 50 $\mu$m size range, but
        it also leaves us with a large and unknown sampling bias. No other study has targeted CAIs in OCs in this size range,
        and consequently we do not know wheter the obtained size distribution and areal density of small CAI fragments are representative for OCs as a
        meteorite class.
        In contrast, for larger CAIs, the results are more certain. The \rev{larger area scanned,} covering three different slabs originating from NWA 5697, gives a smaller
        sampling bias \rev{for this chondrite}. \rev{The uncertainty is still large, given that} we only found one large CAI, \rev{but} the areal density is
        comparable to what has been asserted in the literature for the areal density of CAIs in OCs \citep{2008M&PS...43.1879H,2011Icar..213..547R}.

\section{Conclusions} \label{sec:concl}
We have explored what can be inferred from a dichotomy in the abundance of refractory grains found in chondrites coming from
parent bodies that accreted in the inner and outer solar system. With the hypothesis that this is a consequence of the gravitational
perturbation of Jupiter and potentially Saturn, we performed long-term numerical modeling of a viscous accretion disk with giant planets.
The gap opening at the planet location and related creation of a local pressure maximum result in a size-dependent filtering of dust grains.
We find that the critical grain size above which dust is accumulated outside Jupiter can be characterized by the viscosity and surface density
in the disk, with a power-law dependence, and an exponent that depends on the planet configuration and the detailed description of the
dust diffusion. \rev{As long as more realistic models including angular momentum transport by outflows, a realistic temperature and density structure,
and possibly migrating giant planets do not change qualitatively the gas distribution, the basic picture will not change. But these effects, together with possibly
three-dimensional models, are outside the scope of this paper and will have to be explored in future works.}
Refractionary CAIs are ideal tracers of this dynamics arriving at Jupiter as part of the general infall of gas and dust
from the envelope, because they are essentially unprocessed since their formation. To quantify the dichotomy, we performed highly
resolved $\mu$-XRF and SEM scans of polished slabs from the inner solar system chondrite NWA 5697.
We found one large CAI in the 53 cm$^2$ $\mu$-XRF scan and three small CAIs in the $4.1 \mathrm{cm}^2$ SEM scan, \rev{significantly increasing} the
known inventory of CAIs in OCs. 
The measurements support a critical grain size of $100\textrm{ -- }300 \,\mu m$, which requires the level of turbulent disk viscosity to
be $\alpha\!\lesssim\!5\times10^{-3}$ and suggests a surface density at Jupiter of $\approx\!\!200\,$g cm$^{-2}$, \rev{under the assumption that
the transport of angular momentum is entirely due to viscous processes}. The comparison of three different planet
configurations lets us conclude that, in the implemented framework, the scenario in which Saturn approaches Jupiter close enough to create
a compound gap requires a high level of fine-tuning, and unreasonable model values to be consistent with the laboratory results.
Finally, comparing the areal density of CAIs in OCs to that in carbonaceous chondrites,
\rev{we estimate that the dust mass in the protosolar disk may have been distributed equally inside and outside of Jupiter's orbit,
but we emphasize that this result comes with a considerable uncertainty.}

\acknowledgements
We thank Leonardo Krapp for providing us with part of the numerical tools to model dust in our simulations and the anonymous referee for helpful comments.
The research leading to these results has received funding from the Independent Research Fund Denmark through a DFF Sapere Aude Starting Grant and grant No.\,DFF  8021-00350B (TH) and through grant No.\,DFF 8021-00400B (MEP), from the European Union's Horizon 2020 research and innovation programme through grant agreement No.\,748544 (PBLL), and from the European Research Council (ERC) under the European Union's Horizon 2020 research and innovation programme through
grant agreement No.\,616027 (MB) and grant agreement No.\,638596 (OG).
The Centre for Star and Planet Formation is financed by the Danish National Research Foundation.

GPU computing nodes funded with a research grant from the Danish Center for Scientific Computing and hosted at the University of Copenhagen
HPC facility were used for carrying out the computer modeling. For the laboratory work we used the QuadLab $\mu$-XRF facility at the Danish Natural
History Museum, supported by the ``Noble Gas Timekeepers and Tracers'' grant from the Villum Foundation. We acknowledge the use and support of
the SEM facility at the Geological Survey of Denmark and Greenland, which was instrumental for identification and follow-up confirmation of the CAIs.

This work has made use of IPython \citep{ipython}, NumPy \citep{numpy}, and Matplotlib \citep{Matplotlib} for data analysis and creating figures.

\bibliographystyle{apj}
\bibliography{references}

\end{document}